\documentclass[trackchanges,manuscript]{aastex631}
\usepackage{CJK}

\shorttitle{tidal features and merger rates of HSC-SSP ETGs}
\shortauthors{Huang and Fan}

\graphicspath{{./}{figures/}}

\begin{document}
\begin{CJK*}{UTF8}{gbsn}

\title{Massive Early-Type Galaxies in the HSC-SSP: Flux Fraction of Tidal Features and Merger Rates}

\correspondingauthor{Qifeng Huang, Lulu Fan}
\email{hqf8102@mail.ustc.edu.cn, llfan@ustc.edu.cn}

\author[0000-0003-2863-9837]{Qifeng Huang (黄齐丰)}

\affiliation{Department of Astronomy, School of Physical Sciences, University of Science and Technology of China, Hefei 230026, China}

\author[0000-0003-4200-4432]{Lulu Fan (范璐璐)}

\affiliation{Department of Astronomy, School of Physical Sciences, University of Science and Technology of China, Hefei 230026, China}
\affiliation{Institute of Deep Space Sciences, Deep Space Exploration Laboratory, Hefei 230026, China}
\affiliation{School of Astronomy and Space Science, University of Science and Technology of China, Hefei 230026, China}

\begin{abstract}

Here we present a statistical study on tidal features around massive early-type galaxies (ETGs). Utilizing the imaging data of the Hyper Suprime-Cam Subaru Strategic Program (HSC-SSP), we measure the flux fraction of tidal features ($f_{\rm tidal}$) in 2649 ETGs with stellar mass $M_*>10^{11}M_{\odot}$ and redshift $0.05<z<0.15$ using automated techniques.
The Wide-layer of HSC-SSP reaches a depth of $\sim 28.5$ mag arcsec$^{-2}$ in $i$-band. Under this surface brightness limit, we find that about 28\% of these galaxies harbor prominent tidal features with $f_{\rm tidal}>1\%$, among which the number of ETGs decreases exponentially with $f_{\rm tidal}$, with a logarithmic slope of $\sim100$. Within the stellar mass range we probe, we note that $f_{\rm tidal}$ increases by a factor of 2 from $M_*\approx10^{11}M_{\odot}$ to $M_*\approx10^{12}M_{\odot}$. We also perform pair-count to estimate the merger rate of these massive ETGs. Combining the merger rates with $f_{\rm tidal}$, we estimate that the typical lifetime of tidal features is $\sim$ 3 Gyr, consistent with previous studies. 
\end{abstract}


\section{Introduction} \label{sec:intro}

In the hierarchical scenario of the Lambda Cold Dark Matter ($\Lambda$CDM) cosmological model, galaxies merge successively and grow with time \citep[e.g.][]{Toomre1977,Naab2009,Dokkum2010,RodriguezGomez2016}. Coalescence of galaxies can lead to various phenomena such as active galactic nucleus (AGN) activities, starbursts, and disturbed kinematics of the progenitor galaxies \citep[e.g.][]{Hopkins2008,Hopkins2010,Cappellari2016}. 

Hydro-dynamical simulations have shown that massive galaxies ($M_*\gtrsim 10^{11}M_{\odot}$) assemble mainly through mergers in addition to \textit{in-situ} star formation \citep[e.g.][]{RodriguezGomez2016}, which is also supported by observations \citep[e.g.][]{Dokkum2010,Wel2014}. What's more, the fractional merger rates of galaxies increase steeply with stellar mass \citep[e.g.][]{RodriguezGomez2015,Husko2022}. When spirals merge, the kinetic energy of the rotating disks is transferred into the random motion of stars, thus major mergers or multiple minor mergers play an important role in morphological transformation of galaxies \citep[e.g.][]{Martin2018,Park2021}. These results suggest that massive ETGs with $M_*\gtrsim 10^{11}M_{\odot}$ in the local universe almost certainly have experienced mergers \citep[][]{Stewart2008,Lee2017}, producing tidal features in the process. As a result, we expect to find tidal features with a high frequency around massive ETGs.

Substructures of galaxies such as tidal tails, stellar shells and streams have long been observed. After
years of research, there is little doubt that these features can be produced in galaxy mergers \citep[e.g.][]{Toomre1972,Hood2018,Mancillas2019}. Therefore, tidal features can be used to constrain the merger histories of galaxies, although the inference is often not straightforward. For example, the flux fraction and color of tidal features can be transformed into a lower limit on mass ratio of the progenitors \citep{Gu2013}. Numerical methods have the ability to recover the formation history of the descendant by running a set of simulations and mimicing its morphology \citep{Bilek2022}. Nowadays, with the aid of machine learning techniques and numerical simulations, it is possible to infer the merging histories of galaxies from their integral properties alone \citep{Eisert2022}. If the flux fractions or even the annotated maps of tidal features are used to train the networks, the performance of machine learning may get a pronounced improvement \citep[][the Galaxy Cruise project\footnote{\url{https://galaxycruise.mtk.nao.ac.jp/en/index.html}}]{Sola2022}.

However, most tidal features are extremely faint structures, and their surface brightness fades quickly with time after being produced, making it hard for quantitative measurements. To study the properties of tidal features and their correlations with the host ETGs, a key step is to identify these faint structures from the bright and uneven background. A popular and rather effective way to do this is to inspect the deep images visually \citep[e.g.][]{Atkinson2013,Hood2018,Yoon2020,Bilek2020}, but it is too laborious and time consuming for large samples containing tens of thousands of galaxies. Despite great difficulties, other methods tried to describe tidal features quantitatively and automatically. For example, 2D fitting and decomposition of deep galaxy images has been applied to small samples to extract tidal features from the model-subtracted residual images \citep[e.g.][]{Janowiecki2010,Gu2013,Mantha2019,MartinezDelgado2021}. \citet{KadoFong2018} convolved the images repeatedly and subtracted images with different spatial frequencies to separate tidal features from host galaxy light. Besides, non-parametric programs such as \texttt{NoiseChisel} \citep{Akhlaghi2015a} are able to detect signals deep into the noise. Machine learning methods such as convolutional neural networks are developing fast with a promising future \citep[e.g.][]{Walmsley2019,Hendel2019}. 

In our work, we focus on the flux fraction of tidal features ($f_{\rm tidal}$) and its distribution, and try to establish the connection between flux fraction, visible time of tidal features, and merger rates of their hosts. Simulations have found that the detection rate of tidal features is dependent upon the surface brightness limit and the projection angle \citep[e.g.][]{Ji2014,Mancillas2019,martin2022a}, so what we measured is certainly only the tip of the iceberg of fainter and more extended structures. However, these factors are hard to account for with observational data alone. Future surveys such as the 10-year Legacy Survey of Space and Time (LSST) are required to narrow the gap between observations and reality.

We structure this paper as follows. In Section \ref{sec:data}, we establish the sample of massive ETGs with HSC images based on SDSS spectrum and morphology. And we also present the completeness of the catalog. Section \ref{sec:measure} describes the methods for measuring tidal features in detail, as well as the distribution of $f_{\rm tidal}$ and its correlation with other properties of the host ETG. As tidal features are triggered by mergers, we perform galaxy pair-count to estimate the merger rates of the sample in Section \ref{sec:mr}. In Section \ref{sec:discussion}, we briefly discuss the lifetime of tidal features and the implications of these features on other unobservable quantities of galaxies' past such as the merging histories. Finally, a summary is given in Section \ref{sec:summary}.

Throughout this paper, we use the following cosmological parameters: $H_0 = 70\ \rm{km\ s^{-1} Mpc^{-1}},\ \rm{\Omega}_M = 0.3$ and $\rm \Omega_{\Lambda} = 0.7$. All magnitudes are given in the AB system \citep{Oke1983}.

\section{Data and Sample Selection} \label{sec:data}

\subsection{HSC-SSP Imaging}
The images used in this paper were obtained from the Hyper Suprime-Cam Subaru Strategic Program \citep[HSC-SSP;][]{Aihara2018}. We use the co-added images from the Public Data Release 3 \citep[PDR3;][]{Aihara2022}, which covers over 1000 $\rm deg^2$ and reaches a depth of $\mu_{i}^{\rm lim} \sim 28.5$ mag arcsec$^{-2}$ in $i$-band in its Wide layer when measuring individual galaxies \citep{Huang2018}.
Its relatively large sky coverage and deepness are beneficial to study the LSB substructures of galaxies statistically out to intermediate redshifts \citep{KadoFong2018}. 
Since the data-processing pipeline for the final products of HSC-SSP uses 128$\times$128 binned pixels\footnote{The pixel scale of the images is 0.168".} to estimate the local sky background, it will cause over-subtraction around very extended objects such as the galaxies in our sample, leading to unreliable photometry of LSB structures in the outskirts. 
So we use the intermediate-state images from the HSC archive with only global sky-subtraction performed. In this mode, the size of the superpixels used to estimate the sky background is 1k$\times$1k, large enough for our sample. We use the $r$-band and $i$-band images to extract tidal features, because the surface brightness limits in $z$-band and $y$-band are shallower, and tidal features are less prominent in $g$-band due to their colors.

\subsection{Basic Properties}
We employ stellar masses and redshifts provided by SDSS Data Release 16 \citep[DR16;][]{Ahumada2020}. Stellar masses are calculated by \citet{Chen2012} by fitting the SDSS spectra, adopting the \citet{Bruzual2003} stellar population synthesis models and the \citet{kroupa2001} initial mass function. Redshifts of these galaxies are also obtained from the SDSS spectra.

 To separate the ETGs from the late-type galaxies, we use the galaxy morphology catalog presented by \citet{Meert2015a}. Briefly, \citet{Meert2015a} performed 2D-fittings on $\sim$ 2$\times 10^7$ spectroscopically selected galaxies selected from SDSS DR7. Then they assigned a T-type to each galaxy after calculating the probabilities of being one of four broad galaxy types (Ell, S0, Sab, or Scd), using the method described in \citet{HuertasCompany2011}. ETGs correspond to those galaxies with T-type smaller than 0.5, including lenticular galaxies and ellipticals.
 
 \begin{deluxetable*}{ccc}
\tablecaption{Different galaxy samples.\label{tab:def}}
\tablewidth{\textwidth}
\tablehead{\colhead{Name} & \colhead{Number} & \colhead{Definition}}
\startdata
parent & 3658 & Massive ETGs at $0.05<z<0.15$ with SDSS spectra and clean HSC images\\
paired/unpaired & 1009/2649 & ETGs with/without companions satisfying $\Delta r<2.5$ mag and $r_{\rm proj}<50$ kpc\\
visual/non-visual & 899/1750 & Unpaired ETGs with/without tidal features under visual inspection\\
\enddata
\tablecomments{By using the word ``paired'',  we only focus on the projection distance and flux ratio. It does not necessarily mean they have physically bounded companions.}
\end{deluxetable*}

\subsection{Sample Selection}\label{subsec:selection}
Firstly, we select galaxies brighter than $r_{\rm SDSS}=17.77$ mag from SDSS DR16 \citep{Ahumada2020} and cross-match them with those brighter than $r_{\rm HSC}$ = 19 mag from HSC-SSP PDR3 with clean\footnote{By saying ``clean," we mean that we excluded those galaxies with a saturated or interpolated center. See \url{https://hsc-release.mtk.nao.ac.jp} for more details.} HSC images in both $r$ and $i$-band. Here different magnitude limits are chosen to compensate for the discrepancy between the photometric data of these two surveys. The difference of these two limits is sufficient to cover all SDSS galaxies with HSC imaging, given that magnitude differences between SDSS and HSC data for large galaxies rarely exceed 0.7 mag, according to \citet{Aihara2018a}.

\begin{figure}
    \centering
    \includegraphics[width=\textwidth]{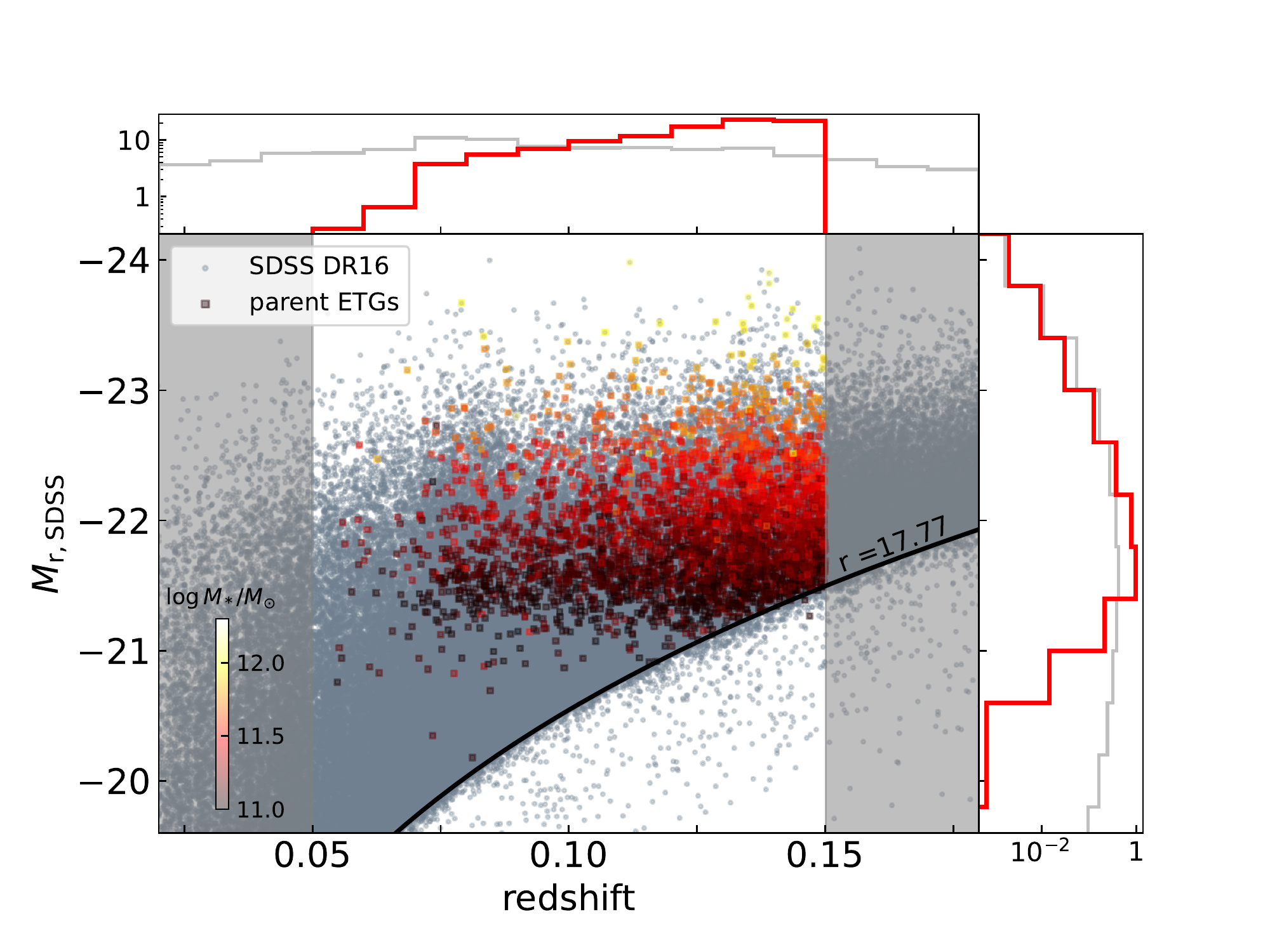}
    \caption{The parent sample used in this paper. Each massive ETG is color-coded by its stellar mass. Gray dots represent all the galaxies with SDSS spectra. The shaded area delineates the redshift cuts of our sample. The histograms of galaxy redshifts and magnitudes are shown in the top and the right plot, respectively. The red lines represent our sample, and the gray lines indicate the SDSS galaxies.}
    \label{fig:1}
\end{figure}

We select for our final sample galaxies in the redshift range $0.05<z<0.15$. At higher redshift, tidal features are hard to observe due to the signal-to-noise ratio of the images and the cosmological surface brightness dimming. What's more, the sample will become more incomplete at higher redshift (Figure \ref{fig:1}). Massive ETGs with redshift lower than 0.05 are rare, and they have much larger angular sizes and less reliable photometric measurements than the galaxies in our sample.

We then limit our sample to massive and early-type galaxies, which refer to galaxies with stellar mass $M_*>10^{11}M_{\odot}$ and T-type $<0.5$, respectively. Finally, we apply a quick visual inspection on our sample and remove those galaxies with obvious spiral patterns or covered by a foreground galaxy, resulting in a sample of 3658 ETGs (Figure \ref{fig:1}). The typical stellar mass-to-light ratio of these ETGs is $3\sim 4 M_{\odot}/L_{\odot}$ in $r$-band.

By cross-matching with the HSC catalog, we find that among these ETGs, 1009 are pair candidates having another galaxy with $\Delta r<2.5$ mag within a projecting distance of 50 kpc. We refer to these galaxies as ``paired ETGs" (Table \ref{tab:def}) and use them only when determining the merger rates in Section \ref{sec:mr}, without measuring their tidal features.
Omitting ``paired ETGs" when measuring tidal features will not introduce any significant bias on the measured distribution of $f_{\rm tidal}$, since we only study tidal features in post-mergers. Besides, it is random to classify post-mergers into ``paired ETGs" mistakenly due to the projection effect.

In the appendix of \citet{Bosch2008}, the authors calculated the mass completeness limit for red galaxies in SDSS as a function of redshifts. At $z=0.15$, the derived limit is $\log M_*/M_{\odot}=10.95$. Thus our sample is complete under this criterion. The high completeness can also be seen in Figure \ref{fig:1}, where only $< 10\%$ of the massive ETGs fall below $r=17.77$ mag. 

However, as the merger rate decreases with cosmic time, it's possible that the lack of faint ETGs at higher redshifts in our sample will force the low-mass ETGs to have less tidal features and less companions. We estimate the impact of this bias as follows. For ETGs with the lowest stellar mass ($\log M_*/M_{\odot}<11.1$) in our sample, the median redshift is 0.118, which increases to 0.132 for an unbiased sub-sample with $\log M_*/M_{\odot}>11.3$. According to cosmological simulations \citep[e.g.,][]{RodriguezGomez2015}, the merger rates at these two redshifts differ by only 2\%. So the redshift evolution of merger rate is too small to affect our results.

Besides, when selecting HSC images, we discard galaxies with a saturate or interpolated center, which will make the photometry less reliable, making our sample biased against ETGs with AGNs or high central surface brightness. Since these properties of galaxies are probably related to their assembly histories \citep[e.g.][]{Hong2015}, this bias should be taken into account when interpreting the results.

\section{Measuring the Tidal features} \label{sec:measure}
\begin{figure}
    \centering
    \includegraphics[width=\textwidth]{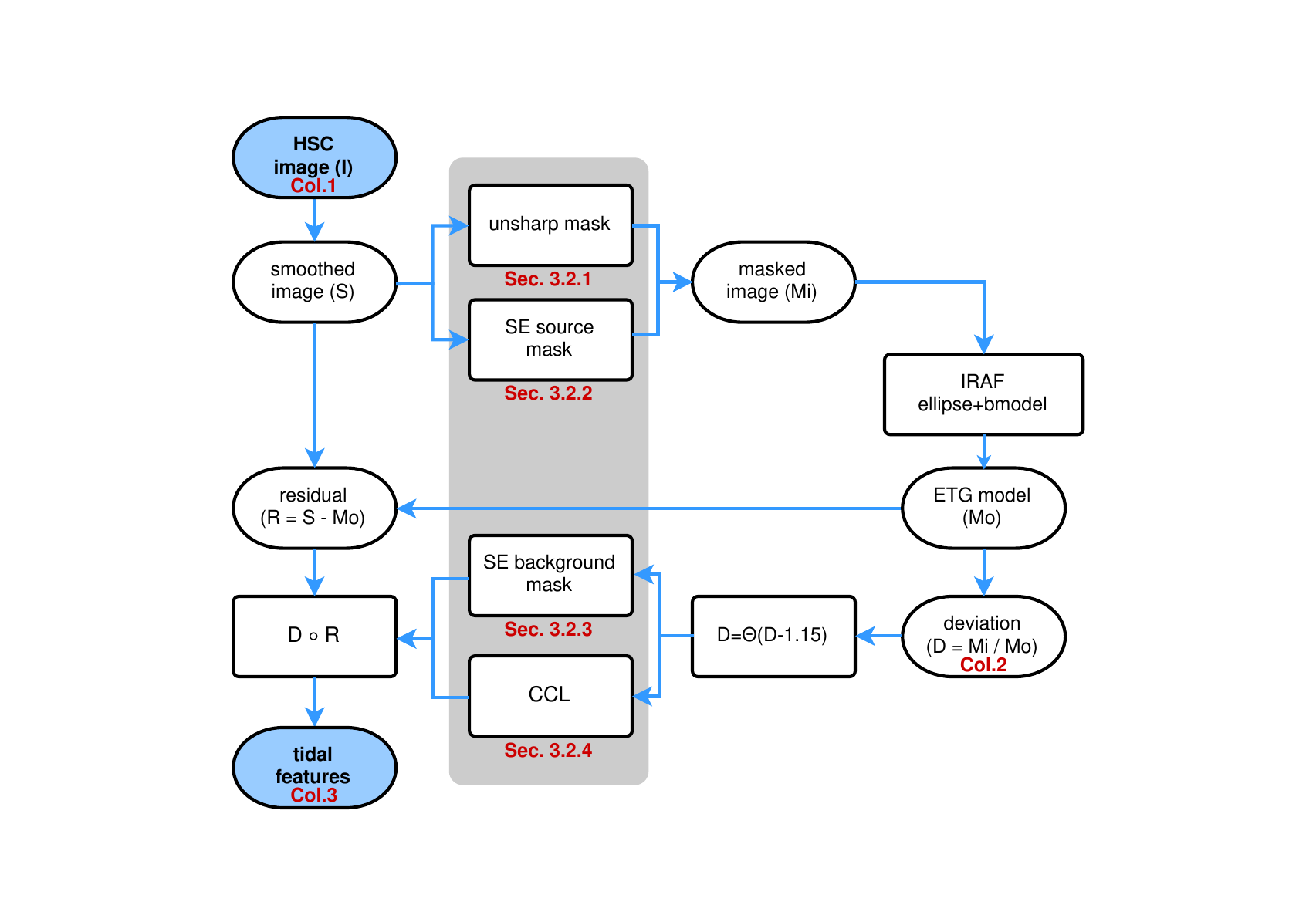}    \caption{Flowchart illustrating the process of extracting and measuring the tidal features. Steps in the shaded box are described in Section \ref{subsec:mask}, while the remaining procedures are explained in Section \ref{subsec:fitting}. Col.1 to Col.3 stand for three columns shown in Figure \ref{fig:4}. Boxes and ovals stand for operations and intermediate results, respectively. The formula $D\circ R$ represents the element-wise multiplication of matrices, and $\Theta(x)$ is the unit step function.}
    \label{fig:2}
\end{figure}

The basic idea of measuring the tidal features is to subtract elliptical isophotal models from the images and analyze the residuals. Up to now, some algorithms have been developed to model the light profile of galaxies, and detect the tidal features automatically \citep[e.g.][]{KadoFong2018,Mantha2019}. In this paper, we mainly refer to the method described in \citet{Dokkum2005} and \citet{Mantha2019} with some adjustments, using \texttt{IRAF} \textit{ellipse} \citep{Jedrzejewski1987} and \textit{bmodel} tasks to build the galaxy models. The whole process is illustrated in a flow chart (Figure \ref{fig:2}).

\subsection{A Quick Visual Inspection}
As mentioned in Section \ref{subsec:selection}, a visual inspection is performed during sample selection to exclude the spirals from our sample. Meanwhile, we also check visually whether these galaxies host tidal features using the $r$-band and $i$-band images, along with colored images online\footnote{\url{https://www.legacysurvey.org/viewer}} if available.

Among the 2649 ETGs in the ``unpaired'' sample, we find that 899 (34\%) of them harbor visible tidal features. However, our visual inspection is crude as there are thousands of images to be examined. There's no doubt that some faint structures, which will become apparent only after some smoothing and contrast adjustment of the images, are missed from our inspection.

\subsection{The Sky Masks} \label{subsec:mask}
\begin{figure}
    \centering
    \includegraphics[width=0.5\textwidth]{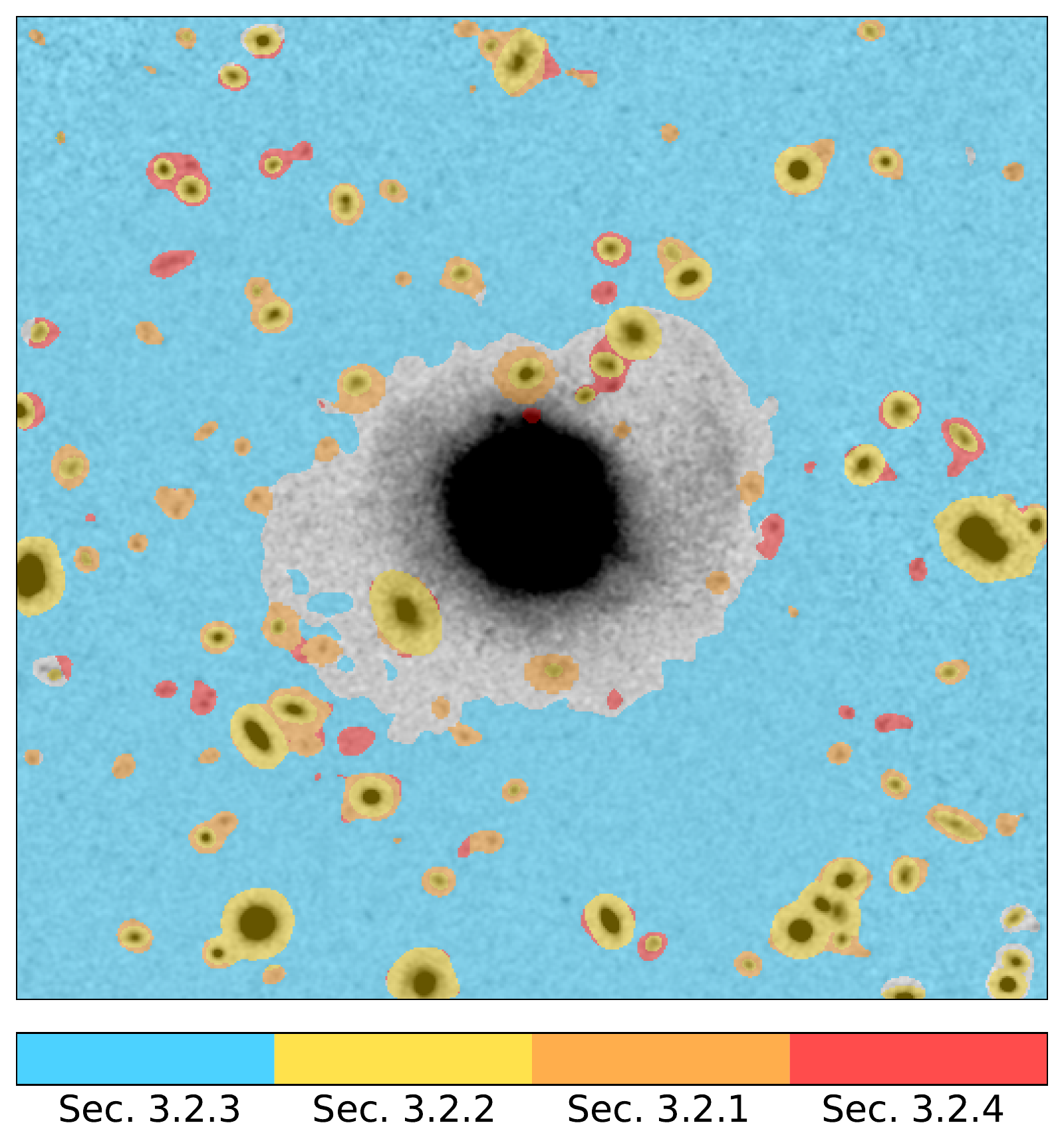}
    \caption{Visualization of four masking steps. The $i$-band negative image of an elliptical is shown in the background. Different masks described in Section \ref{subsec:mask} are marked in different translucent colors. Blue: \texttt{SExtractor} (sky background). Yellow: \texttt{SExtractor} (sources). Orange: unsharp masking. Red: false detection removal through Connected-component Labeling.}
    \label{fig:3}
\end{figure}

The majority of tidal features are extremely faint. Measurements on their flux fractions can be severely affected by other sources with higher surface brightness. So our first step is to detect and mask these sources to avoid contamination. 
Several methods of masking for individual bands are described in the subsections below. And an example of the masking procedures is illustrated in Figure \ref{fig:3}.

For our classical method used here, it's almost impossible to have all the images well-masked with a single set of parameters. A trade-off has to be made between avoiding the noise and missing the signal. 

\subsubsection{Unsharp masking}\label{subsubsec:unsharp}
For most galaxies in the background and stars in the foreground, their angular sizes are significantly smaller than the ETGs in our sample (i.e., with much higher spatial frequencies). Unsharp masking is a common technique to detect these signals. We convolve the image to smooth the noise first, using kernels with different Full Width at Half Maximum (FWHM) for $r$-band and $i$-band images due to different seeing conditions. Then, by comparing the image with a Gaussian-smoothed version of itself, pixels showing a deviation larger than a certain threshold are masked, excluding the center region of the target ETG. Neighboring pixels of the masked pixels are also masked to cover the outer wings. In practice, we find that some sharp tidal features, especially strong tidal tails or shells, may be mistakenly masked through this technique. To mitigate the risk, we require that the masked area should have an ellipticity smaller than 0.7, or it will be abandoned.

\subsubsection{Source Extractor detection}\label{subsubsec:SEx}
For larger sources with size comparable with the target galaxy, or sources with extended outskirts, unsharp masking is insufficient to cover them cleanly. We use \texttt{\texttt{SExtractor}} \citep{Bertin1996} to generate additional masks for these sources. Before running \texttt{SExtractor}, we apply the arcsinh transformation to the images for contrast stretching, increasing the weight of the faint outskirts during source detection. The configuration parameters of \texttt{SExtractor} are shown in Table \ref{tab:SEx}.

\texttt{SExtractor} derives the shape parameters of sources by measuring their isophotal profiles. We draw ellipses based on these parameters to act as masks and enlarge them according to the magnitudes of the corresponding sources, with larger magnification applied to brighter ones. Again, tidal features with high surface brightness are sometimes misidentified as separate sources due to incorrect segmentation, which has always been a challenge for algorithms using the imaging data alone. These segmentation errors may lead to underestimation of $f_{\rm tidal}$, which is one of the main uncertainties in our analysis.

\begin{deluxetable}{lcc}
\tablecaption{Configuration parameters of Source Extractor for masking sources (Section \ref{subsubsec:SEx}) and  the sky background (Section \ref{subsubsec:bsm}).\label{tab:SEx}}
\tablewidth{\textwidth}
\tablehead{\colhead{Parameter} &  \colhead{Source} & \colhead{Background} }
\startdata
DETECT\_MINAREA & 5 & 10\\
DETECT\_THREASH & 2.0 & 2.0\\
DEBLEND\_MINCONT & 0.01 & 0.05\\
BACK\_SIZE & 16 & 512
\enddata
\end{deluxetable}

\subsubsection{Background masks} \label{subsubsec:bsm}
In deep images, LSB structures are often mixed with noise, leading to significant false detection if a low threshold is chosen to retain these structures. By estimating and masking the sky background, the contamination of noise can be greatly reduced. The segmentation maps provided by the HSC pipeline \citep{Bosch2018} are not optimized for LSB structures with a detection limit of only $\mu_i^{\rm lim} \sim 25.5$ mag arcsec$^{-2}$. To make better use of the depth of HSC images, we run \texttt{SExtractor} again with a different set of parameters (Table \ref{tab:SEx}) and mask the background using the resultant segmentation maps. 
The borders of these optimized sky masks approximately follow the isophote of $\mu_{i}\ga 27$ mag arcsec$^{-2}$, much deeper than the original one. A detailed discussion on the surface brightness limit is presented in Section \ref{subsec:sb}.

\subsubsection{Connected-component labeling} \label{subsubsec:CCL}
To remove the undetected sources that miss the detection of unsharp masking and \texttt{SExtractor}, and to remove clustered noise, we perform Connected-component Labelling (CCL) with 8-connectivity on the residuals using \texttt{SciPy} \citep{Virtanen2020}. Under the assumption that tidal features are relatively faint and extended structures, we remove those components with a coverage less than 500 pixels. 
This threshold is empirically chosen based on the PSF FWHM and the convolution kernel to match the typical size of faint and point-like sources while leaving the tidal features less affected. 


\subsection{Fitting and Extraction\label{subsec:fitting}}
We use \texttt{IRAF} \textit{ellipse} task to fit the light profiles of ETGs through the smoothed images after applying the masks described in Section \ref{subsubsec:unsharp} and \ref{subsubsec:SEx}. Then we generate elliptical galaxy models using the \textit{bmodel} task. After the galaxy models are built, we apply the remaining masks described in Section \ref{subsubsec:bsm} and \ref{subsubsec:CCL} and compare the smoothed images with the models (Figure \ref{fig:2}). Areas in the smoothed and masked images with flux 15\% larger than the corresponding models are identified as tidal features. The threshold is set to ensure consistency with visual inspection and with the choice in \citet{Dokkum2005} for a comparison.
We allow the ellipticity and the position angle to vary freely with radius. The maximum wandering of the central point between consecutive isophotes is limited to avoid unexpected fatal errors. And we don't perform the masking and fitting processes iteratively since it may affect model fitting by only masking the pixels on the right tail of the noise distribution, which will underestimate the model flux.

\begin{figure*}
    \centering
     \includegraphics[width=\textwidth]{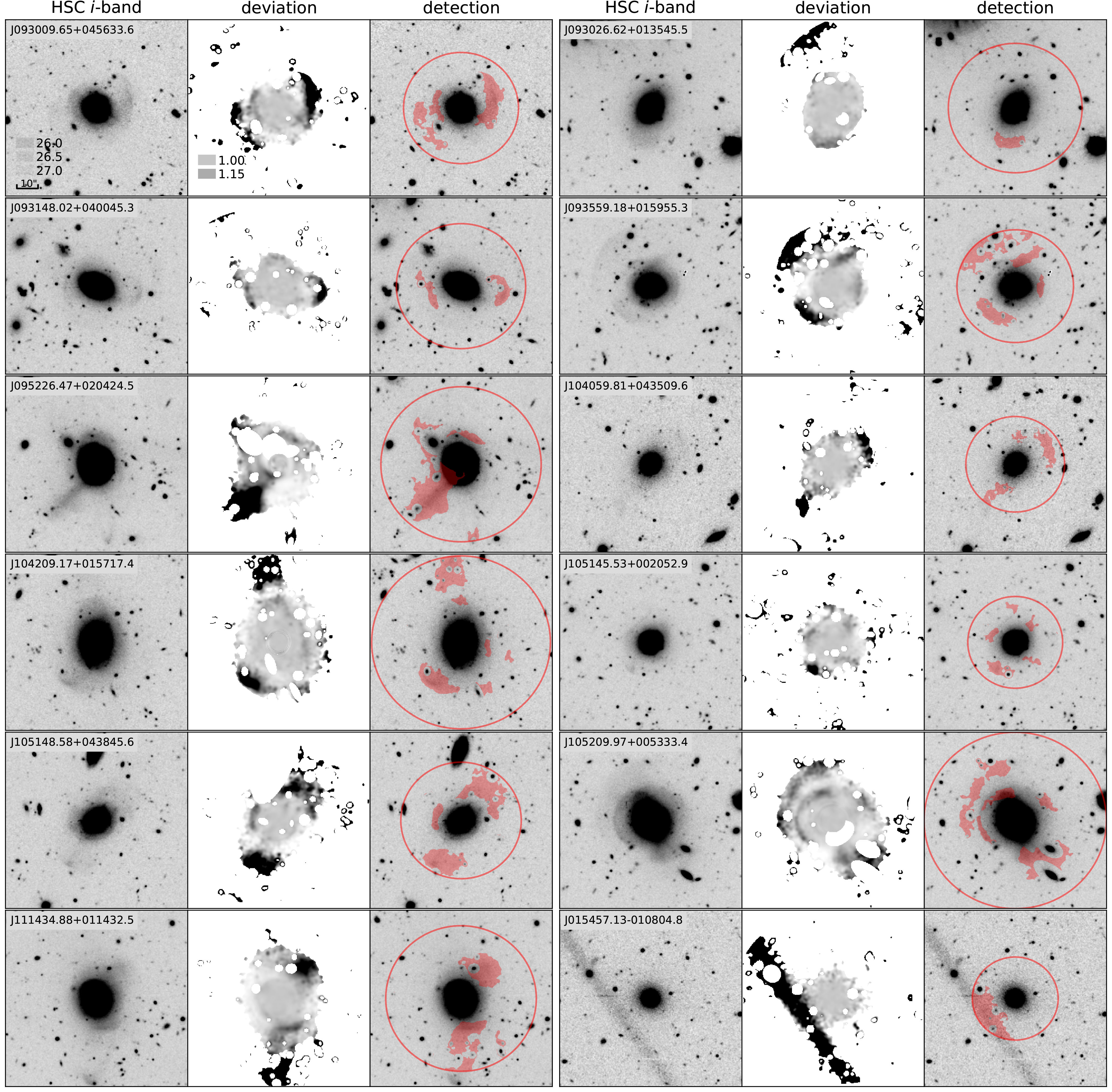}
    \caption{Twelve examples of extracting tidal features. The left column shows the smoothed $i$-band HSC images. The middle column comes from the quotient of the smoothed images and the models. Darker color represents larger deviation from the models. The right column marks the detected features in red, with the maximum radius for detection ($2.5R_{25,i}$) labeled by the red circle. All the images are in the same scale as indicated in the first example. The last example presents the contamination of artifacts.}
    \label{fig:4}
\end{figure*}

We present some examples of the detection results in Figure \ref{fig:4}.  The original images downloaded from the HSC database are shown in the first column. The middle column comes from the quotient of the smoothed images and the models. All masks described in Section \ref{subsec:mask} are plotted in white. The third column is identical to the first one, with detected tidal features in red color overlaid. In the last example (the bottom-right panel), we show that artifacts can lead to false detection.

After running all the images and plotting the radial distribution of detected pixels, we find that almost all of the features detected lie within $2.5R_{25,i}$, 2.5 times the radius where surface brightness of the host ETGs drop to 25 mag arcsec$^{-2}$ in $i$-band, while pixels detected outside $2.5R_{25,i}$ are dominated by contamination such as unmasked wings of other sources and artifacts. So we limit our detection range to $2.5R_{25,i}$, as indicated with the red circles in Figure \ref{fig:4}. For a galaxy with $i=17$ mag and an $R^{1/4}$-profile, this radius corresponds to 15 times the effective radius. At $2.5R_{25,i}$, the surface brightness drops to ${\mu_i}>27$ mag arcsec$^{-2}$ for the majority of ETGs in our sample.

Finally, the flux of tidal features is calculated by summing up the flux of the detected pixels in the residual (Figure \ref{fig:2}).

\subsection{Results}\label{subsec:result}
\subsubsection{Flux fraction and its distribution}\label{subsubsec:fluxfrac}
As mentioned in Section \ref{subsec:selection}, we perform tidal feature measurements on the unpaired sub-sample of 2649 ETGs without a notable companion, which refers to the satellites with flux ratio greater than 1:10, to focus on the tidal features of post-mergers. The measured magnitudes of tidal features and their host ETGs are available online\footnote {\url{https://github.com/llfan-ustc/huang2022/blob/main/catalog.csv}. Examples and descriptions of the data are given in Appendix \ref{sec:supdata}.}.

In Figure \ref{fig:5}, we show the distribution of flux fraction of tidal features ($f_{\rm tidal}$) of these massive ETGs, which is defined as the flux ratio of the tidal features to the whole galaxy:
\begin{equation}
    f_{\rm tidal}\equiv\frac{F_{\rm tidal}}{F_{\rm total}}\approx\frac{F_{\rm tidal}}{F_{\rm tidal}+F_{\rm model}}\label{eq:1}
\end{equation}
Here we use the sum of flux of tidal features and flux of the isophotal model to substitute for the total flux.
Since we choose tidal features with 15\% flux excess, $F_{\rm total}$ calculate here may actually slightly smaller than its true value. However, the magnitudes calculated using $F_{\rm total}$ are consistent with SDSS\footnote{Magnitudes of extended objects given by HSC-SSP PDR3 may suffer from excessive background subtraction. In $i$-band, on average, they are 0.11 mag fainter than SDSS photometry and 0.16 mag fainter than our results.}, supporting that the difference is negligible. 

On average, $f_{\rm tidal}$ measured in $r$-band image is 0.10 dex larger than that measured in $i$-band, implying that tidal features are $\sim0.25$ mag bluer than their hosts in the $r-i$ color. This can be seen from Figure \ref{fig:5} that the gray histogram exceeds the blue one. However, as tidal features are extracted independently from the images of these two bands, the detected regions and shapes of the galaxy models are not necessarily the same. So large uncertainties on the color may exist. 

For further analysis, we mainly use the mean values of $f_{\rm tidal}$ measured in $r$-band and $i$-band since their distributions are similar, and random errors can be reduced by averaging. We use the scatter of $f_{\rm tidal}$ in these two bands as an estimate of the typical uncertainty, assuming that the intrinsic scatter result from the color of tidal features and their hosts is relatively small.

According to whether there are tidal features observed through the quick visual inspection, we divide the ``unpaired'' sample into two parts, namely ``visual'' and ``non-visual''. The solid histograms in Figure \ref{fig:5} represent all 2649 ETGs and the hollow ones correspond to ETGs with visually discernible tidal features, i.e., the ``visual'' sample. The consistency between these quantitative measurements and qualitative classification is apparent --- the majority of ETGs with large $f_{\rm tidal}$ are labeled as ``visual'' and vise versa. However, there are still discrepancies on a level of $\sim15\%$. The main reasons for this are that faint but extended tidal features sometimes escape our quick visual inspection, artifacts and LSB features are hard to distinguish, and bright stars or galaxies in the images can cause contamination. Particularly, bright nearby sources, artifacts and fatal errors of \texttt{IRAF} \textit{ellipse} encountered in the fitting processes can often lead to false detection with very large $f_{\rm tidal}$, as indicated by the drop of the green line shown in Figure \ref{fig:5}.

Measuring ETGs showing a hardly disturbed morphology often results in uncertainties larger than the average. Unmasked contamination, especially clustered noise which is hard to separate from faint tidal features, makes the data with $f_{\rm tidal}\la 0.5\%$ rather noisy. So the distribution of $f_{\rm tidal}$ at the small end offers little useful information apart from its proportion. Because of this, along with the consistency level with visual inspection mentioned above, we define an ETG to have prominent tidal features if $f_{\rm tidal}>1\%$. And the fraction of ETGs with $f_{\rm tidal} > 1\%$ is denoted as $f_{1\%}$.

Among these 2649 ETGs, we see that 1909 (72\%) of them have $f_{\rm tidal}<1\%$. For the remaining ETGs with prominent tidal features, number of ETGs in each bin decreases monotonously, which can be well described by an exponential distribution (except for a few outliers at the high-$f_{\rm tidal}$ end):

\begin{equation}
    N(x<f_{\rm tidal}<x+{\rm d} x)\sim e^{-\alpha x}{\rm d} x,\ {\rm for}\ f_{\rm tidal}\geq 0.01,\label{eq:2}
\end{equation}
where the variable $x$ is in the decimal form. For the reasons mentioned above, we can characterize the distribution of $f_{\rm tidal}$ using two parameters: $f_{\rm 1\%}$ and the logarithmic slope $\alpha$. To obtain $\alpha$, we fit the data with $f_{\rm tidal}>1\%$ using the Maximum A Posteriori estimation (Appendix \ref{sec:mae}). The best-fit parameters are shown in Table \ref{tab:para}. Roughly speaking, the number of massive ETGs decreases by a factor of $e$ when $f_{\rm tidal}$ increases by every 1\%. Notice that the fitting is performed on the ``unpaired'' sample. The result of visual inspection (the ``visual'' sample) is only used for comparison.

\begin{deluxetable}{cccc}
\tablecaption{Proportion of unpaired ETGs with prominent tidal features ($f_{1\%}$) and the best-fit parameters of Equation \ref{eq:2}.\label{tab:para}}
\tablewidth{\textwidth}
\tablehead{Sample & \colhead{$f_{\rm 1\%}$} & \colhead{$\alpha$}}
\startdata
all unpaired                   & $27.9\%\pm1.0\%$ & $100.8^{+3.8}_{-3.6}$\\
$11.0<\log M_*/M_{\odot}\leq11.2$ & $24.1\%\pm3.5\%$ & $106.1_{-5.7}^{+6.5}$\\
$11.2<\log M_*/M_{\odot}\leq11.4$ & $28.6\%\pm2.9\%$ & $95.6_{-5.8}^{+6.5}$\\
$\log M_*/M_{\odot}>11.4$      & $36.4\%\pm3.4\%$ & $99.8_{-6.6}^{+7.6}$
\enddata
\end{deluxetable}

\begin{figure}
    \centering
    \includegraphics[width=\textwidth]{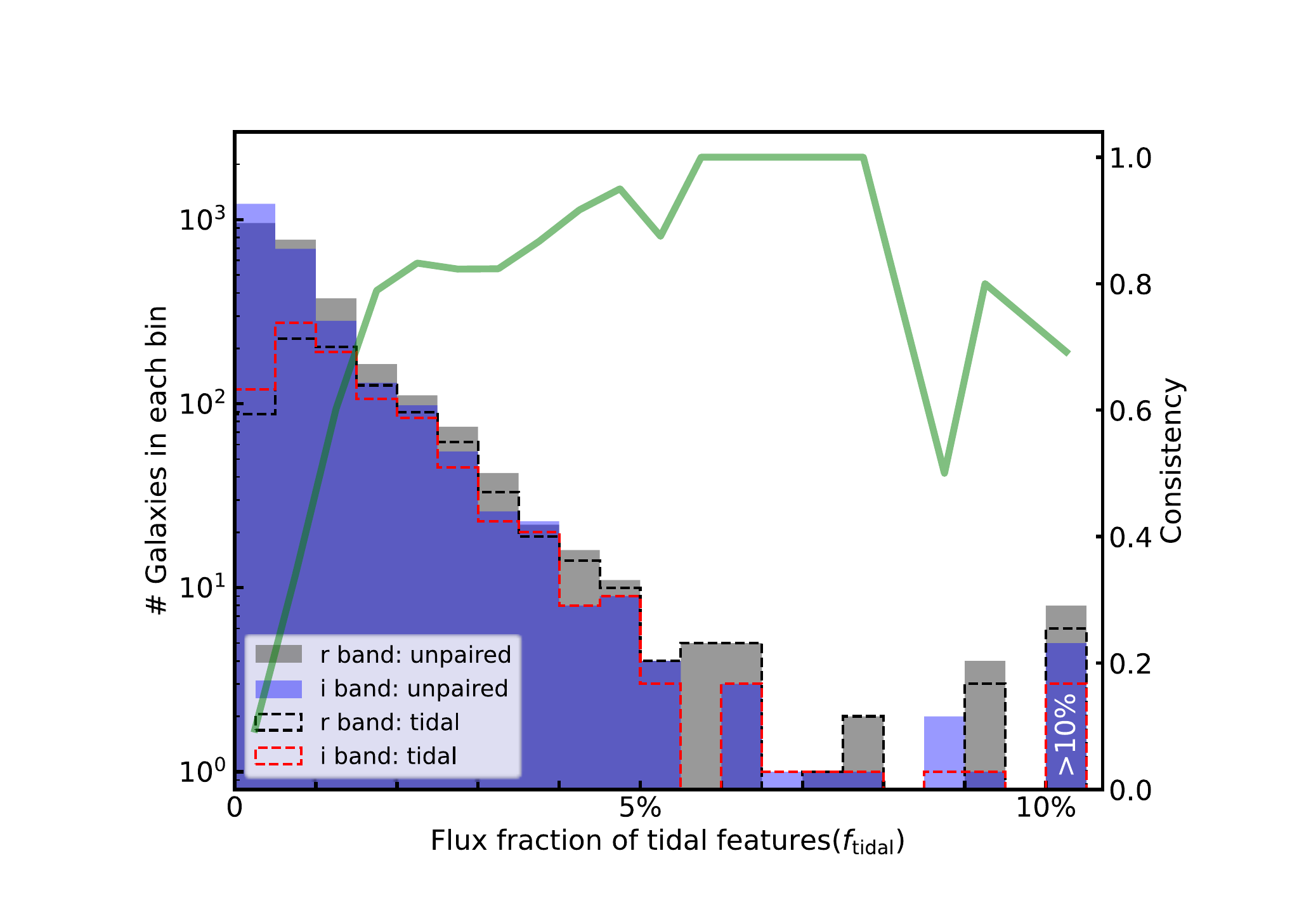}
    \caption{Distribution of flux fraction of tidal features ($f_{\rm tidal}$) in massive ETGs. The solid histograms represent the ``unpaired" ETGs, while the hollow one represents the ``visual'' sample. Plots in blue and gray stand for data in $i$-band and $r$-band, respectively. The green line is obtained by dividing the number of ``visual'' ETGs and ``unpaired" ETGs in each bin, which is a rough estimate of the consistency between quantitative measurements and our quick visual inspection. Ideally, it will rise steeply and remain near 1.0.}
    \label{fig:5}
\end{figure}

\subsubsection{Tidal features and host stellar mass}\label{subsubsec:corr}

Based on previous studies, both the merger rates and the \textit{ex-situ} mass fraction of ETGs have been found to increase with stellar mass, which implies that tidal features are likely to appear more frequently in more massive ETGs. In both observations and simulations, visual inspections have shown that the fraction of ETGs showing tidal features indeed increases with stellar mass \citep[e.g.][]{Bilek2020,Yoon2020,martin2022a}. For example, \citet{Yoon2020} found that only $2\%\sim5\%$ of ETGs with dynamical mass $M_{\rm dyn}< 10^{10.4}M_{\odot}$ show tidal features in coadded images of SDSS Stripe 82, while this fraction increases to about 30\% to 40\% for massive ETGs with $M_{\rm dyn}> 10^{11.4}M_{\odot}$. 

\begin{figure*}
    \centering
    \includegraphics[width=\textwidth]{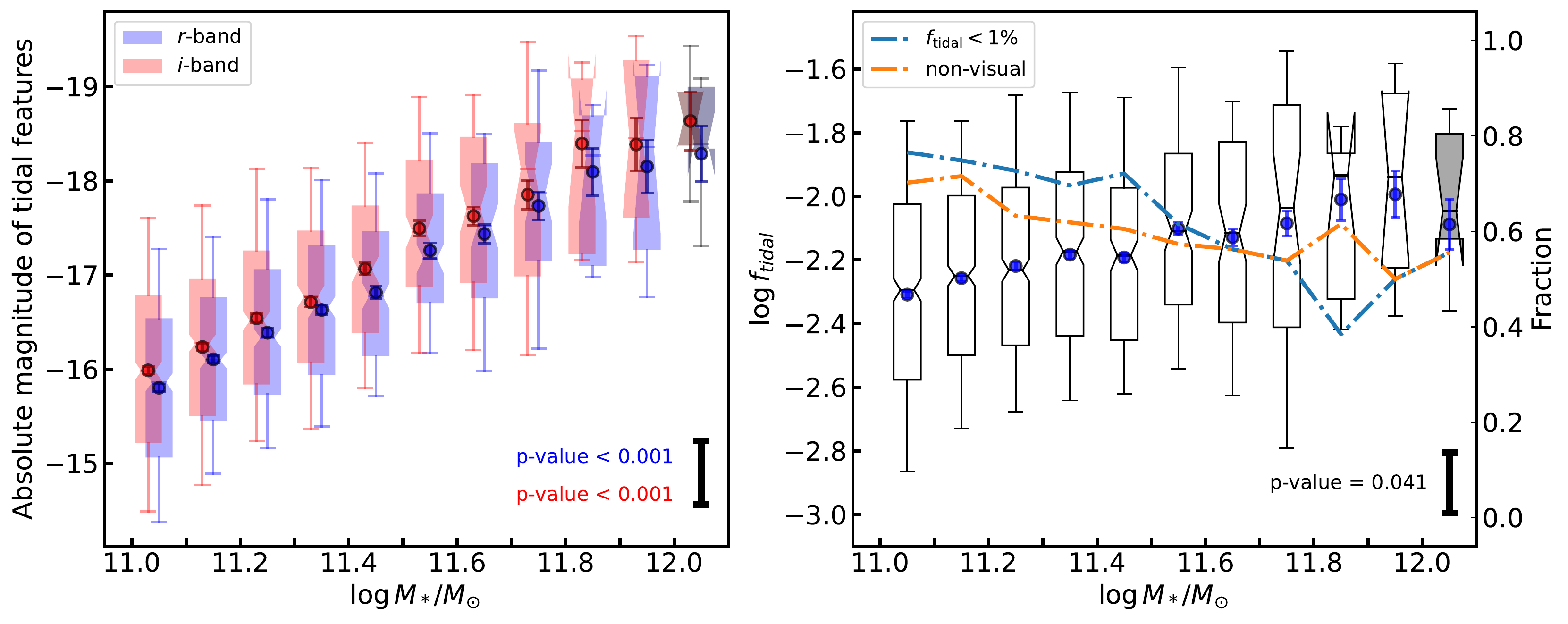}
    \caption{{\bf Left:} Boxplots of absolute magnitude of tidal features against host stellar mass, color-coded for the values measured in $r$-band (blue) and $i$-band (red), respectively. Each box represents the median magnitude value (small dashes) and extends from the 25\% to the 75\% quartiles. The lower and upper bars encompass between 10\% and 90\% of the distributions, respectively. The dots with error bars represent the mean values of each box. The right-most boxes in darker colors contain all ETGs with $\log M_*/M_{\odot}>12.0$. The black error bar at the bottom right indicates the typical error for single galaxies. And p-values of the Pearson correlation test for testing non-correlation are also shown. {\bf Right:} Boxplot of flux fractions of tidal features ($f_{\rm tidal}$) against host stellar mass. All legends are identical to those used in the left panel. The dashed lines show the proportion of ETGs with $f_{\rm tidal}<1\%$ and ETGs belong to the ``non-visual'' sample} in each bin.
    \label{fig:6}
\end{figure*}

For the unpaired ETGs in our sample with quantitative tidal feature measurements, we plot the absolute magnitudes and the flux fraction of detected features against stellar mass of their host ETGs in Figure \ref{fig:6}. From the left panel, we can see that the \textit{luminosity} of tidal features increases steadily with stellar mass. What's more, as shown in the right panel, we find that the \textit{flux fraction} of tidal features also increases with stellar mass. Within the mass range we probe, the median value of $f_{\rm tidal}$ increases by a factor of 2, from about $0.5\%$ at $M_*\approx 10^{11.0}M_{\odot}$ to $1\%$ at $M_*\approx 10^{12.0}M_{\odot}$. The drop seen at $M_*> 10^{12}M_{\odot}$ may be simply an artifact since our image cutouts are not large enough to cover the detection range defined in Section \ref{subsec:fitting} for a few of the most massive galaxies in our sample.

\begin{figure*}
    \centering
    \includegraphics[width=\textwidth]{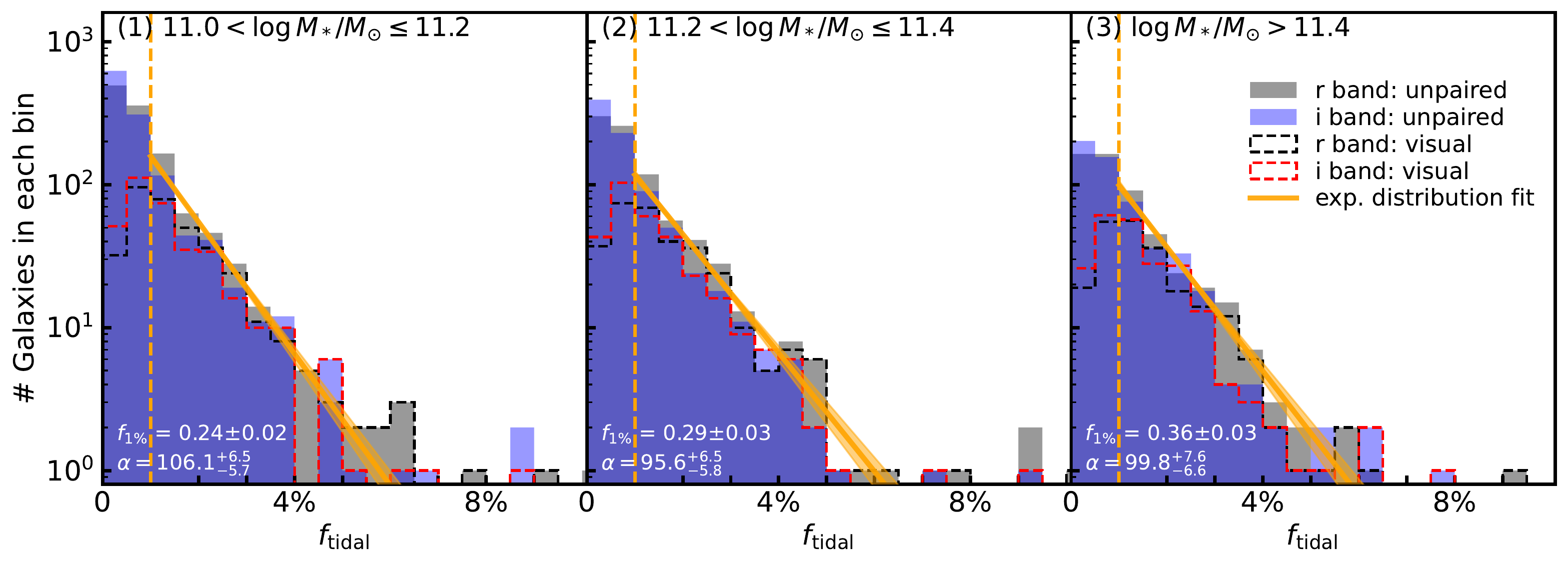}
    \caption{Distributions of flux fraction of tidal features ($f_{\rm tidal}$) of ETGs in different stellar mass bins. The left, centre, and right panels correspond to stellar mass range of $10^{11.0-11.2}, 10^{11.2-11.4}$, and $>10^{11.4}M_{\odot}$, respectively. The legends are identical to those in Figure \ref{fig:5}. Fractions of ETGs with $f_{\rm tidal}>1\%$ and the best fitting parameters $\alpha$ are presented in the bottom left of each panel.}
    \label{fig:7}
\end{figure*}

We further examine how the slope $\alpha$ changes with stellar mass. As shown in Figure \ref{fig:7}, the best fitting parameters for massive ETGs with $f_{\rm tidal}>1\%$ do not show, if any, significant differences among three stellar mass bins. A wider mass range extended to well below $10^{11}M_{\odot}$ is necessary to verify this constancy.

The physical implications of these results will be briefly discussed in Section \ref{subsec:lifetime}.

\section{Pair Count and Merger Rates} \label{sec:mr}
Galaxy mergers play a crucial role in the evolution of massive ETGs, and they are the origin of tidal features (Section \ref{sec:intro}). In recent years, intensive researches on galaxy merger rates have been carried out based on counting galaxy pairs \citep[e.g.][]{Lotz2011,Man2016,Mantha2018} or using numerical simulations \citep[e.g.][]{RodriguezGomez2015,OLeary2021,Husko2022}. Here we perform pair-count on our ETG sample in Section \ref{subsec:pairfrac}, estimate the merging timescales to calculate the merger rates in Section \ref{subsec:mt} and compare the results with some previous works in Section \ref{subsec:compare}.

If the fraction of galaxy with companions and the typical merging timescale is known, the fractional merger rate, the number of merger events per galaxy per unit time, can be estimated as the quotient of these two quantities \citep[e.g.][]{Lotz2011,Conselice2014,Mantha2018}:
\begin{equation}
    R_{\rm merg}=\frac{f_{\rm pair}}{\langle T_{\rm merg}\rangle} \label{eq:3}
\end{equation}
where $f_{\rm pair}$ stands for the proportion of paired systems under certain selection criteria, $\langle T_{\rm merg}\rangle$ represents the mean observable timescale of these galaxy pairs. 

In some cases, there is more than one satellite around the central ETGs. Considering that merger events are highly independent with each other, we calculate the pair fraction by using the number of companions rather than paired ETGs:
\begin{equation}
    f_{\rm pair}=N_{\rm companion}/N_{\rm ETG}\label{eq:4}
\end{equation}
Hereafter, we adopt the convention that major and minor mergers (or companions) correspond to flux ratio limits of 1:1--1:4 and 1:4--1:10 in $r$-band, respectively.

\subsection{Close Pair Fraction}\label{subsec:pairfrac}

\begin{figure}
    \centering
    \includegraphics[width=0.8\textwidth]{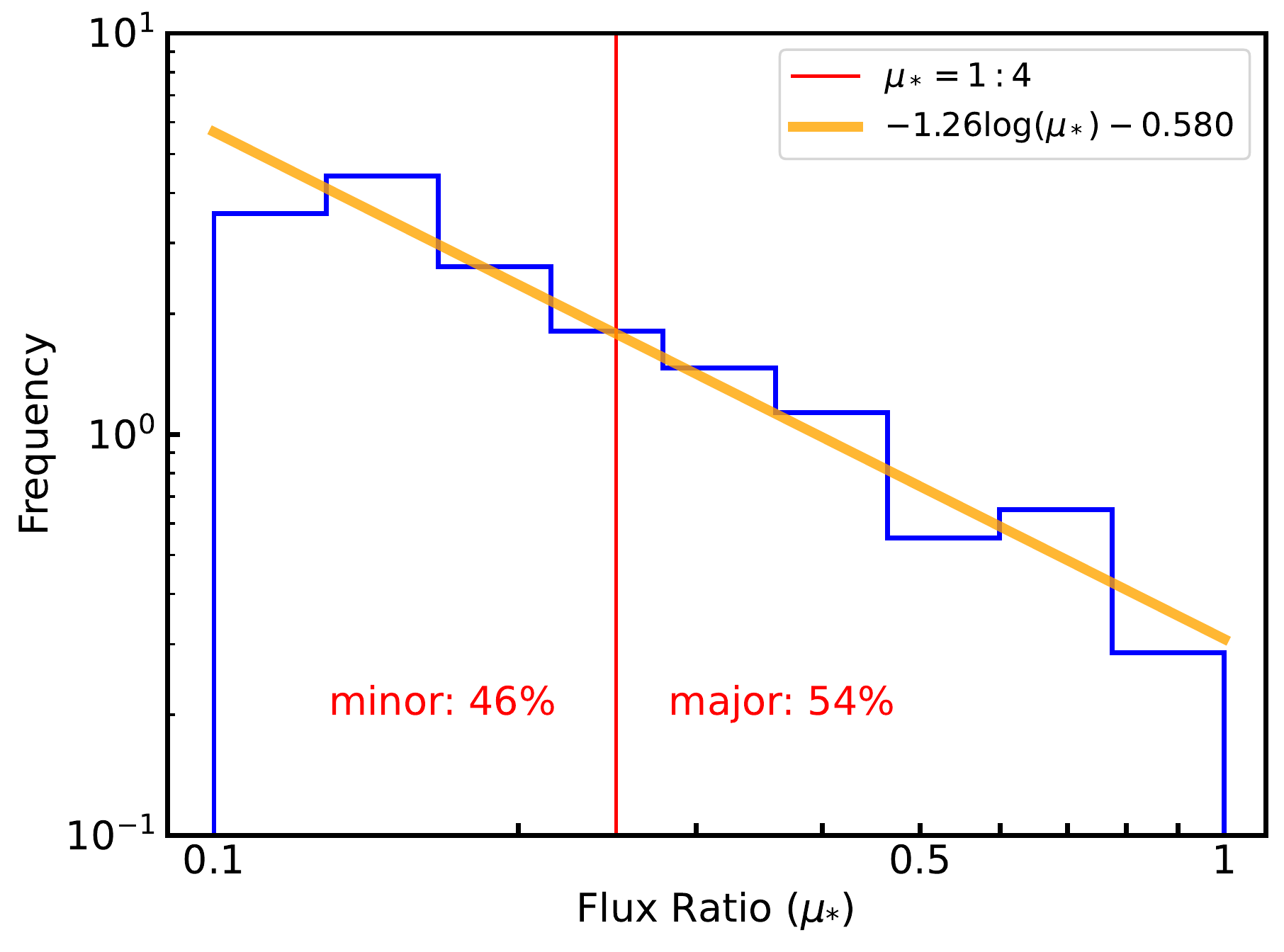}
    \caption{Distribution of the flux ratio between the companions and corresponding hosts. The vertical line separates major mergers from minor mergers. And the orange solid line shows the best exponential fit to the data.}
    \label{fig:8}
\end{figure}

In observations, galaxy pairs are defined as two galaxies with similar redshifts and within a given projection distance ($r_{\rm proj}$) range. The pair fraction directly depends on the range of $r_{\rm proj}$ used to select galaxy pairs. To be consistent with the literature, we use different $r_{\rm proj}$ ranges to calculate the pair fraction and merger rates. The commonly used ranges include 5--30 kpc, 5--50 kpc, 14--43 kpc, 7--28 kpc \citep[e.g.,][]{Lotz2011,Newman2012,Man2016,Mundy2017,Husko2022}. As the first step, we search for galaxies pairs in the HSC catalog with $r_{\rm proj}<50$ kpc and a flux ratio in $r$-band larger than 1:10 ($\Delta r<2.5$ mag) as the candidates, as described in Section \ref{subsec:selection}. Although the stellar mass ratios are not always available, \citet{Mantha2018} demonstrated that the use of flux ratios and mass ratios gives highly consistent results at redshift $z<1$.

In the line of sight, galaxies with radial velocity difference $\Delta v\leq 500$km/s are likely to be gravitationally bound \citep[e.g.][]{Patton2000}, and this criterion is commonly used in previous work \citep[e.g.][]{Patton2008,Tasca2014,Mantha2018}. Spectroscopic redshifts given by SDSS DR16 \citep{Ahumada2020} and Galaxy And Mass Assembly (GAMA) survey DR3 \citep{Baldry2018} are used to cross-match with our candidates mentioned above. When a spectroscopic measurement is lacking, which is the case for over 80\% of the candidates, we use the photometric redshifts (photo-$z$) from SDSS DR16 and the DESI Legacy Imaging Surveys \citep{Zhou2021}, which covers all the galaxies with $z<21$ mag in the fields of HSC-SSP. If the spectroscopic redshift of the host ETG lies within the 1$\sigma$ error of a candidate's photo-$z$, then this candidate is considered as a physically bound companion.
By visually inspecting the images, we find that the fluxes of some sources near the central ETGs are obviously overestimated by the HSC catalog, probably due to poor background estimation or segmentation errors. We also find that these sources lack reliable photometric redshifts, with no matched redshift data or with $\sigma_{\rm phot} \gtrsim 0.06(1+z_{\rm phot})$, where $z_{\rm phot}$ and $\sigma_{\rm phot}$ stand for the photo-$z$ and its error, respectively. So we remove these sources. A different threshold on the photo-$z$ errors has little impact on the results.

We summarize our selection criteria below:
\begin{eqnarray}
    &1.\ |&z_{\rm spec,ETG}-z_{\rm spec,c}| < 500\ {\rm km/s} \nonumber\\
    & {\rm or} \nonumber\\ 
    &2.\ |&z_{\rm spec,ETG}-z_{\rm phot,c}| <  \sigma_{\rm phot,c};\ \sigma_{\rm phot,c} < 0.06(1+ z_{\rm phot,c}) \label{eq:5}
\end{eqnarray}
where the term in absolute value is the redshift difference between the host ETG and the companion, the subscript ``c" refers to the companions.

By applying the criteria listed in Equation \ref{eq:5}, we select 593 companions around 545 ETGs from the parent sample. Fractions of ETGs with major or minor companions are listed in Table \ref{tab:mfmr}. The distribution of the flux ratio between the companions and hosts is shown in Figure \ref{fig:8}.  We note that the numbers of major mergers and minor mergers are comparable for our sample, consistent with previous observations and cosmological simulations in the stellar mass range probed in our work. \citep[e.g.][]{Man2016,Husko2022}.

\begin{deluxetable*}{cCCCCC}
\tablecaption{Merger Fractions and Merger Rates\label{tab:mfmr}}
\tablewidth{2\textwidth}
\tablehead{$r_{\rm proj}$ range & \multicolumn2c{Pair fraction} & $\langle T_{\rm merg}\rangle$ & \multicolumn2c{$R_{\rm merg}$} \\
(kpc) & {} & {} & ($\rm Gyr$) &  \multicolumn2c{($\rm Gyr^{-1}$)}\\
\cline{2-3}\cline{5-6}
\  & $\rm major$ & $\rm major+minor$ & \  & $\rm major$ & $\rm major+minor$
}
\startdata
7--28  & {0.036\pm 0.003} & {0.064\pm 0.004} & {0.85\pm0.08} & {0.043\pm 0.006} & {0.075\pm 0.009}\\
14--43 & {0.053\pm 0.004} & {0.103\pm 0.005} & {1.19\pm0.12} & {0.045\pm 0.005} & {0.086\pm 0.010}\\
5--30 & {0.042\pm 0.003} & {0.076\pm 0.004} & {0.99\pm0.10} & {0.042\pm 0.005} & {0.077\pm 0.009}\\
5--50 & {0.083\pm 0.005} & {0.153\pm 0.006} & {1.79\pm0.18} & {0.046\pm 0.005} & {0.086\pm 0.009} 
\enddata
\end{deluxetable*}

\subsection{Merging Timescale}\label{subsec:mt}

Different methods and definitions have been used in the literature to determine the merging (observable) timescales of galaxy pairs. For example, \citet{Lotz2011} and \citet{Kitzbichler2008} determined the merging timescales with the aid of numerical simulations. \citet{Patton2000}, \citet{Dokkum2005} and \citet{Tal2009} used the dynamical friction timescales described in \citet{Binney1987}, which often gives a good rough estimation for merging timescales \citep{Conselice2014}, while some other works assumed constant timescales \citep[e.g.][]{Man2016}. In this paper, we choose to use the Equation 7 of \citet{Jiang2014}, which is based on the dynamical friction timescales calibrated to simulations.

For each paired system in our sample, the observable timescale for the pair within the annular $r_1<r_{\rm proj}<r_2$ is estimated as follows:

\begin{equation}
T_{\rm merg}=0.892\ \frac{M_{\rm 1,vir}}{M_{\rm 2,vir}}\left[M_{\rm 1,vir}GH_0E(z)\right]^{-1/3}(r_{\rm 2}-r_{\rm 1}) \label{eq:6}
\end{equation}
where $E(z)=\Omega_{\Lambda}+\Omega_{\rm M}(1+z)^3$ and $G$ is the gravitational constant. $M_{\rm 1,v}$ and $M_{\rm 2,v}$ stand for the median virial masses of isolated halos which host galaxies with stellar masses of the host ETGs and the companions, respectively. The accuracy of Eqation \ref{eq:6} is at 10\% level \citep{Jiang2014}. To convert the stellar mass to the halo mass, we use the result for red galaxies provided by \citet{Velander2014}:

\begin{equation}
M_{\rm vir}=1.43\times10^{13}h_{70}^{-1}M_{\odot}\left(\frac{M_*}{2\times10^{11}h_{70}^{-2}M_{\odot}}\right)^{1.36}, \label{eq:7}
\end{equation}
in which $h_{70}=1$ for the Hubble constant used in this paper.

By combining Equation \ref{eq:6} and \ref{eq:7}, we calculate the mean merging timescales under different pair selection criteria. With the pair fraction in hand, we can then derive merger rates of these massive ETGs in our sample by applying Equation \ref{eq:3}. The results are shown in Table \ref{tab:mfmr} and a comparison with previous work is given in the following section.

\subsection{Comparison with Simulations}\label{subsec:compare}

Despite intense investigations, merger rates obtained from past surveys and simulations do not agree well. Here we compare our results with those from some simulations The ratio limits of major and minor mergers mentioned below are the same as our work, except that they used stellar mass ratios rather than flux ratios. Using the Illustris simulation, \citet{RodriguezGomez2015} presented a fitting function for the dependence of galaxy-galaxy merger rate on redshift, stellar mass, and mass ratio. Choosing $z=0.1214,M_*=2.75\times10^{11}M_{\odot}$, which are the mean values of our sample\footnote{We calculate the mass of the descendants in our sample by adding the flux of the pairs and assuming that their mass-to-light ratios are the same.}, it gives a major merger rate of $0.060\ \rm{Gyr^{-1}}$ and a minor+major merger rate of $0.118\ \rm{Gyr^{-1}}$.
\citet{OLeary2021} used the EMERGE simulation to explore the galaxy-–galaxy merger rate. They found that the major merger rate of massive galaxies (the stellar mass of the main progenitor larger than $10^{11.0}M_{\odot}$) by selecting pairs in the projection distance range 14-43 kpc is $0.069\pm 0.005 {\rm\ Gyr^{-1}}$.
\citet{Husko2022} studied the statistics of galaxy mergers using the GALFORM semi-analytical model of galaxy formation \citep[e.g.][]{Cole2000,Baugh2019}.
According to Figure 5 in their paper, for galaxies at $z\approx 0.12$ with the same stellar mass range of our sample, the major merger rate is $0.055{\rm\ Gyr^{-1}}$.

We see that the merger rates shown in Table \ref{tab:mfmr} are overall smaller than those given by simulations , which is common when performing such comparisons \citep[e.g.,][]{Robaina2010,Man2016,Mundy2017}, although the opposite result also exists \citep[e.g.,][]{Man2012}.

\section{Discussion} \label{sec:discussion}

\subsection{Surface Brightness Limit}\label{subsec:sb}

\begin{figure}
    \centering
    \includegraphics[width=0.8\textwidth]{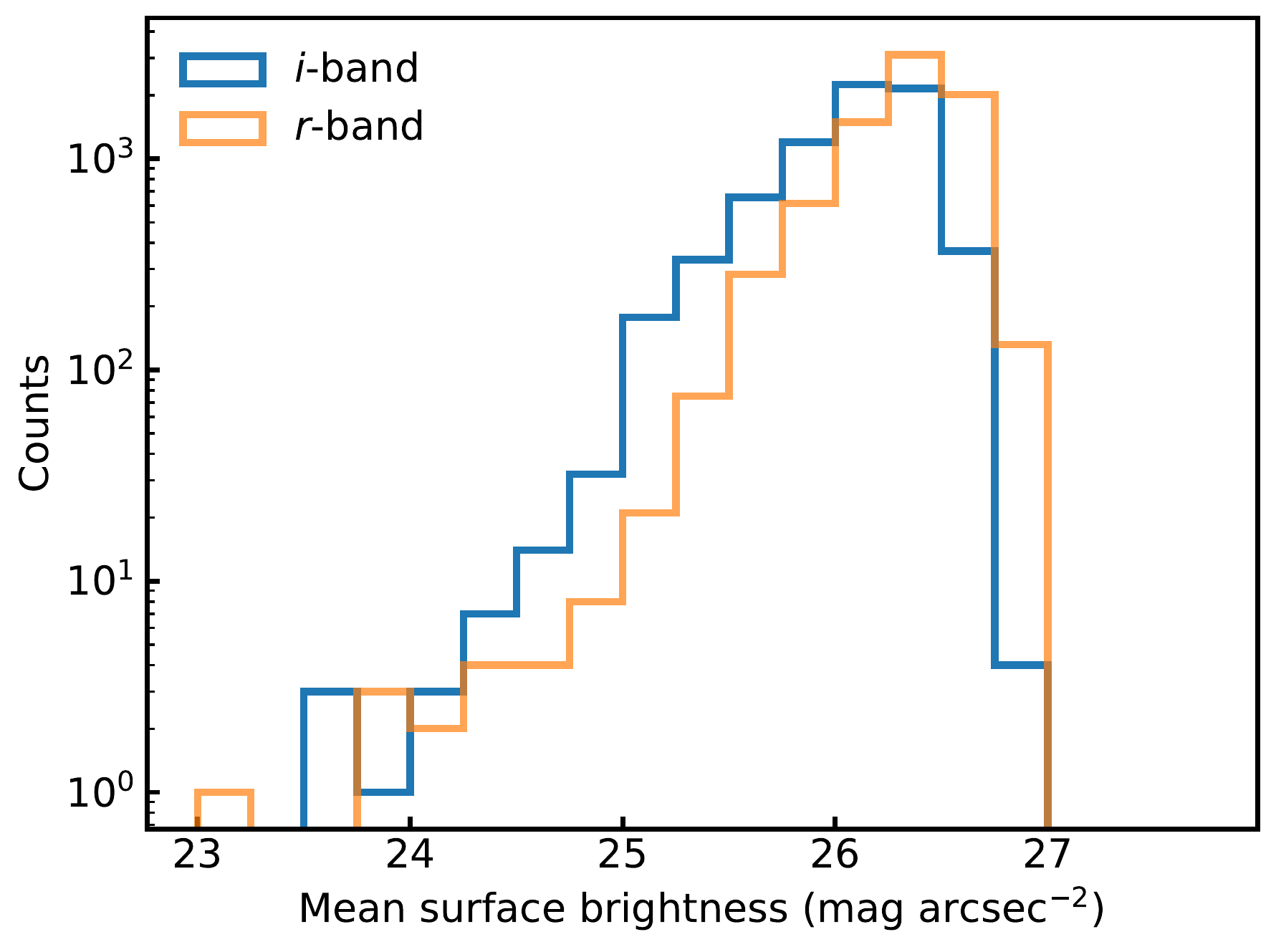}
    \caption{Histogram of the mean surface brightness values for connected regions detected in $i$-band and $r$-band.}
    \label{fig:9}
\end{figure}

The observability of tidal features depends strongly on the surface brightness limit \citep[e.g.][]{Ji2014,Mancillas2019}. And numerical simulations have shown that the majority of the tidal features are visible only when the limiting surface brightness goes down to $\gtrsim$ 30 mag arcsec$^{-2}$ in the $r$-band \citep[e.g.][]{Johnston2008,martin2022a}. In Figure \ref{fig:9}, we present the distribution of the average surface brightness of all the connected areas detected by the algorithm, showing a detection limit of $\mu_{r,i}^{\rm lim}\approx 27$ mag arcsec$^{-2}$. The difference between $i$-band and $r$-band is probably due to different noise levels between these two bands and the intrinsic color of tidal features. Compared with \citet{Sola2022}, there seems to be so few tidal features with high surface brightness ($\mu_{r,i}\lesssim 25$ mag arcsec$^{-2}$) in our sample. The discrepancy in bright tidal features could be due to the following reasons: 1) Ongoing mergers are excluded from our sample, in which bright features often occur. 2) Incorrect segmentation of \texttt{SExtractor} leads to masks on these features. 3) \citet{Sola2022} did not perform model subtraction before measuring the surface brightness of tidal features.

As shown in the literature \citep[e.g.][]{KadoFong2018,Sola2022}, in observations, the surface-brightness limit for reliable tidal feature detection is always much shallower than the nominal depth of the images. This difference is due to various reasons, such as the apparent size of tidal features and methods used to estimate these limits. In our work, another factor contributing to this discrepancy is the mask we use in Section \ref{subsubsec:bsm} to exclude clustered noise from the detection. Also, a significant fraction of tidal features is expected to extend to a galactocentric radius beyond $25R_{\rm eff}$ for the most massive galaxies \citep{martin2022a}, which is much larger than our choice ($\sim 15R_{\rm eff}$) in Section \ref{subsec:fitting}. The fact that we can not detect tidal features at large galactocentric radii is also caused by the surface brightness limit, since tidal features gradually dissipate and become fainter as they move outwards.

\subsection{Lifetime of Tidal Features}\label{subsec:lifetime}

\begin{figure*}
    \centering
    \includegraphics[width=0.7\textwidth]{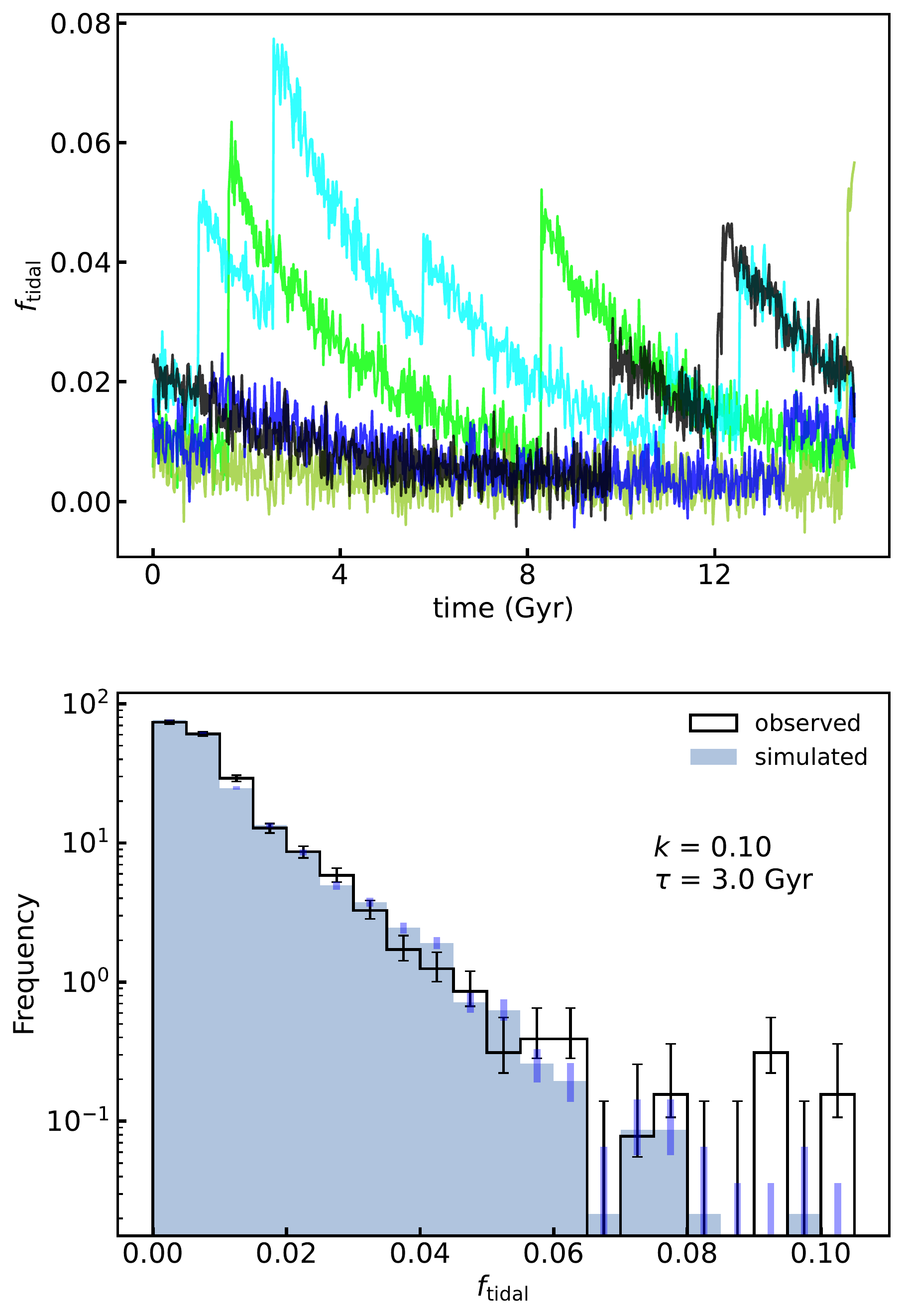}
    \caption{A toy model on the production of tidal features. {\bf Top:} five random examples of how $f_{\rm tidal}$ evolves according to Equation \ref{eq:9}. In this model, tidal features are created instantaneously and then decay exponentially. {\bf Bottom:} a comparison of the observed distribution of $f_{\rm tidal}$ (mean value of $i$-band and $r$-band) and a simulated distribution with $k=0.10,\ \tau=3.0$ Gyr. Error bars represent the Poisson errors.}
    \label{fig:10}
\end{figure*}

To better understand the frequency that tidal features occur, the visible timescales of tidal features should be taken into account. But so far, a detailed study on the evolution of LSB features is still lacking. Here we attempt to estimate the lifetime of tidal features through two independent approaches based on our data, where the surface brightness limit is $\mu_{r,i}^{\rm lim}\approx$ 27 mag arcsec$^{-2}$ (Section \ref{subsec:sb}).

As we can only observe a snapshot of the state of galaxies, a widely used method is to infer the timescales from proportions. From Section \ref{subsubsec:fluxfrac} and Section \ref{subsec:pairfrac}, we see that the fraction of physically paired ETGs in the parent sample ($16.0\%\pm0.7\%$) is comparable with ETGs that have prominent tidal features in the unpaired sample ($27.9\%\pm1.0\%$). As a rough estimate, this implies that the lifetime of tidal features ($t_{\rm tidal}$) is approximately 1.7 times the corresponding merging timescale, resulting in $t_{\rm tidal}\sim$ 3 Gyr.\footnote{Actually, we are dealing with two different populations. For pair-count, $M_*=10^{11}M_{\odot}$ is the lower limit of the \textit{progenitor} mass, while for tidal feature detection, it is the lower limit of the \textit{descendent} mass. However, we have checked that this discrepancy has little effect on our results in this section.}

Intuitively, if these features fade too fast, the distribution of $f_{\rm tidal}$ is likely to be steeper than now, and $f_{\rm 1\%}$ may become impossibly small. So another possible way to estimate the lifetime is by analyzing its influence on the shape of $f_{\rm tidal}$ distribution (i.e., the slope $\alpha$ and $f_{\rm 1\%}$).

The initial value of $f_{\rm tidal}$ immediately after the final coalescence\footnote{More precisely, it is the beginning of the period when only one core is visible in the merging system.} (defined as $t=0$) depends on many factors, such as mass ratio, gas fraction, orbital parameters of the progenitors, surface brightness limit of the observations and viewing angle. But the influences of these factors may be extremely complicated and are wildly uncertain without dedicated simulations. Here we make a simplified assumption that the flux of tidal features at $t=0$ is proportional to that of the original satellite, and $f_{\rm tidal}$ decreases exponentially\footnote{Since there is no previous study on the detailed evolution of the luminosity of tidal features, we have tried several analytical models, such as linear functions and power laws. Among these, the exponential model has only one parameter that describes the characteristic timescale and it works well.} over time with the characteristic timescale $\tau$:
\begin{equation}
    f_{\rm tidal}(t) = \frac{k\mu_*}{1+\mu_*}e^{-t/\tau}\label{eq:9}
\end{equation}
Here $k$ is a scale factor representing the flux ratio of the original satellite and resultant tidal features at $t=0$, which we assumed as constant under a certain surface brightness limit. And $\mu_*$ stands for the flux ratio of the progenitors.

We simulate a set of 10000 galaxies and let them evolve for 15 Gyr following Equation \ref{eq:9}, and add random Gaussian noise $f_{\rm tidal}$ to mimic ``mini mergers'' with flux ratio smaller than 1:10 and false detection caused by clustered noise. 
Five examples are shown in the top panel of Figure \ref{fig:10}, and the bottom panel shows the distribution of $f_{\rm tidal}$ at the last epoch. 
At each time step, a galaxy has a chance to merge with a satellite and produces tidal features. The probability of merging is determined by the merger rates, which is set to 0.08 Gyr$^{-1}$ here following Section \ref{sec:mr}. And flux ratio of the progenitors ($\mu_*$) is sampled from the fitting function of the observational data shown in Figure \ref{fig:8}. 
These processes will result in a distribution of $f_{\rm tidal}$ similar to the observed one, and its characteristic parameters $f_{\rm 1\%}$ and $\alpha$ depend on the scale factor $k$ and the timescale $\tau$ in equation \ref{eq:9}. The most significant difference is that the number of observed ETGs does not vanish at the high $f_{\rm tidal}$ end, which can be explained by contamination and that some mergers act more violently than average. 
Such a long duration of evolution mentioned above is unlikely in the real universe, but here the distribution of $f_{\rm tidal}$ needs time to reach equilibrium from the randomly set initial conditions. 

\begin{figure}
    \centering
    \includegraphics[width=\textwidth]{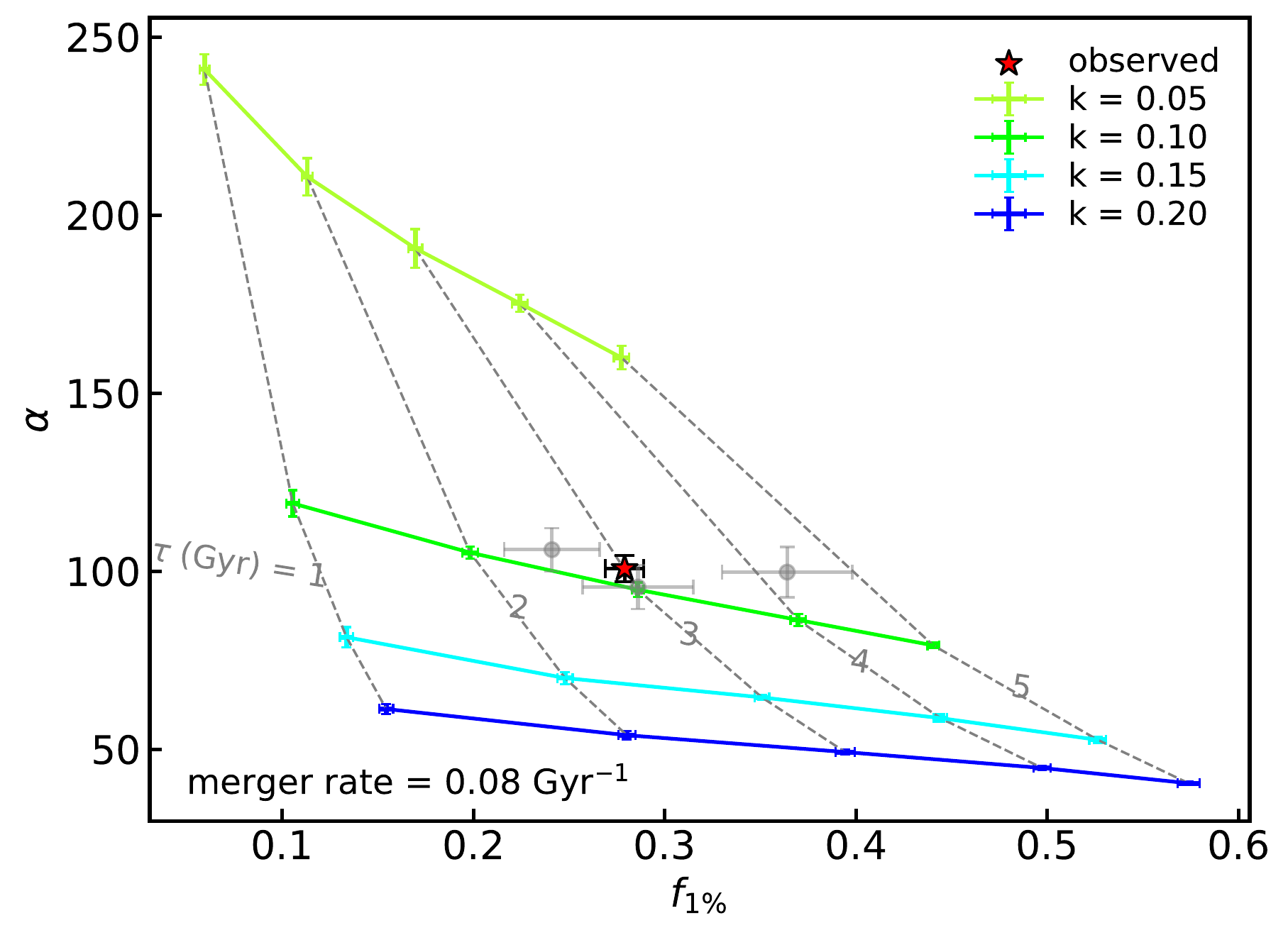}
    \caption{The dependence of $f_{\rm tidal}$ distribution on the parameters of our toy model, $k$ and $\tau$. Each line represents a value of $k$. The five points on it, starting from the left, correspond to $\tau({\rm Gyr})=$ 1, 2, 3, 4, 5, or $\tau\times{\rm merger\ rate}=$ 0.08, 0.16, 0.24, 0.32, 0.40, respectively. The locations of these 20 points are obtained through simulation rather than lengthy analytical calculations, so error bars are attached to them. The observed value of ($f_{\rm 1\%},\alpha$) is marked with a star. Its position illustrates that $k\approx 0.10$ and $\tau\approx$ 3 Gyr. The three gray dots represent the observed values corresponding to different mass bins defined in Section \ref{subsubsec:fluxfrac}.}
    \label{fig:11}
\end{figure}

We plot the dependence of $f_{\rm 1\%}$ and $\alpha$ on $k$ and $\tau$ in Figure \ref{fig:11} by simulating 200 Monte Carlo randomized sets of galaxies. Placing the observed values $f_{\rm 1\%}=27.9\%\pm1.0\%$ and $\alpha=100.8_{-3.6}^{+3.8}$ in it, as indicated with the red star, we see that $k\approx 0.10$ and the lifetime $\tau\approx$ 3 Gyr, which fit the observed distribution of $f_{\rm tidal}$ well. Incidentally, $k\approx 0.10$ implies that $\sim$ 10\% of the stars in the satellites are transformed into tidal features with surface brightness $\mu_{r,i}\la27$ mag arcsec$^{-2}$ immediately after merging. The gray dots in Figure \ref{fig:11} represent ETGs with different stellar mass, as shown in Table \ref{tab:para}. No significant difference in the value of $k$ is found. The discrepancy in the horizontal direction is more likely due to the difference in merger rates rather than a significant change in the lifetime of tidal features.

Although the two methods are crude, they give similar estimations on $t_{\rm tidal}$, and they are consistent with previous numerical simulations and observations \citep{Ji2014,Mancillas2019,Yoon2020}.

\section{Summary} \label{sec:summary} 

In this paper, we measure the flux fraction of tidal features ($f_{\rm tidal}$) in massive ETGs ($M_*>10^{11}M_{\odot}$) quantitatively on a statistical level, using $r$-band and $i$-band images of 2649 sources with $0.05<z<0.15$ in the Wide layer of HSC-SSP PDR3. The \texttt{IRAF} \textit{ellipse} task is used to fit the light profiles of these ETGs, and irrelevant sources are carefully masked to reduce contamination. We investigate how $f_{\rm tidal}$ correlates with stellar mass in the high mass regime. We also calculate the merger rates of massive ETGs based on counting close pairs. By combining the merger rates and the distribution of $f_{\rm tidal}$, we give an estimate of the lifetime of tidal features under the surface brightness limit of the Wide layer of HSC-SSP. Our main conclusions are summarised as follows.

\begin{itemize}
    \item For massive ETGs ($M_*>10^{11}M_{\odot}$) with prominent tidal features ($f_{\rm tidal}\ga 1\%$), the number of ETGs decreases roughly exponentially with $f_{\rm tidal}$, with a logarithmic slope of $100.8_{-3.6}^{+3.8}$. No significant variation on this result is found within the stellar mass range probed here.
    \item On average, both the luminosity of tidal features and $f_{\rm tidal}$ increase monotonously with stellar mass of the host ETGs. The median value of $f_{\rm tidal}$ increases from about $0.5\%$ to $1\%$ when stellar mass increases from $M_*\approx10^{11}M_{\odot}$ to $M_*\approx10^{12}M_{\odot}$.
    \item We provide the merger rates ($R_{\rm merg}$) of these massive ETGs by counting galaxy pairs and assuming the dynamical friction timescales as the merging timescales. The merger rate for major or minor mergers is about $0.08\ {\rm Gyr^{-1}}$. Detailed results are listed in Table \ref{tab:mfmr}.
    \item By applying a toy model to describe the observed distribution of $f_{\rm tidal}$, we find that for these massive ETGs, the lifetime of tidal features is $\sim $ 3 Gyr under the surface brightness limit of $\mu_{r,i}^{\rm lim}\approx$ 27 mag arcsec$^{-2}$, which is consistent with the result given by comparing the number of galaxy pairs and the number of ETGs with prominent tidal features. Since the lifetime of tidal features depends on their morphology (e.g., shells have a longer lifetime than tails. \citet{Mancillas2019}), the value given here may act as an average. In dry mergers where star formation is negligible, at least $\sim$ 10\% of the stars in the satellites are transformed into tidal features immediately after merging.
\end{itemize}

The possibility of inferring merging histories based on observed tidal features at present time is alluring. To perform this on large samples, better extraction algorithms, observational data and numerical simulations are required for future studies. What's more, tidal features with different morphology differ in origin and lifetime, so it's better to study them separately. In this sense, automatic methods that can divide the detected features into different categories will be beneficial \citep[e.g.][]{Hendel2019}.

\begin{acknowledgments}

We thank the anonymous referee for constructive comments and suggestions. We also thank Prof. Xu Kong, Mr. Guangwen Chen and Mr. Bojun Tao for helpful discussions. LF gratefully acknowledges the support of the National Natural Science Foundation of China (NSFC, grant No. 12173037), the China Manned Space Project with NO. CMS-CSST-2021-A04, Cyrus Chun Ying Tang Foundations and the Strategic Priority Research Program of Chinese Academy of Sciences, Grant No. XDB 41010105.

The Hyper Suprime-Cam (HSC) collaboration includes the astronomical communities of Japan and Taiwan, and Princeton University. The HSC instrumentation and software were developed by the National Astronomical Observatory of Japan (NAOJ), the Kavli Institute for the Physics and Mathematics of the Universe (Kavli IPMU), the University of Tokyo, the High Energy Accelerator Research Organization (KEK), the Academia Sinica Institute for Astronomy and Astrophysics in Taiwan (ASIAA), and Princeton University. Funding was contributed by the FIRST program from the Japanese Cabinet Office, the Ministry of Education, Culture, Sports, Science and Technology (MEXT), the Japan Society for the Promotion of Science (JSPS), Japan Science and Technology Agency (JST), the Toray Science Foundation, NAOJ, Kavli IPMU, KEK, ASIAA, and Princeton University. 
This paper makes use of software developed for Vera C. Rubin Observatory. We thank the Rubin Observatory for making their code available as free software at http://pipelines.lsst.io/.
This paper is based on data collected at the Subaru Telescope and retrieved from the HSC data archive system, which is operated by the Subaru Telescope and Astronomy Data Center (ADC) at NAOJ. Data analysis was in part carried out with the cooperation of Center for Computational Astrophysics (CfCA), NAOJ. We are honored and grateful for the opportunity of observing the Universe from Maunakea, which has the cultural, historical and natural significance in Hawaii.

Funding for the Sloan Digital Sky Survey IV has been provided by the Alfred P. Sloan Foundation, the U.S. Department of Energy Office of Science, and the Participating Institutions. 
SDSS-IV acknowledges support and resources from the Center for High Performance Computing at the University of Utah. The SDSS website is www.sdss.org.

\end{acknowledgments}
\appendix

\section{Supplementary Data} \label{sec:supdata}

Here we provide a catalog of 2649 massive ETGs in our parent sample with tidal features measurement. In Table \ref{tab:tidal}, we present the first 15 sources and the illustrations of each column. The full catalog is available at \url{ https://github.com/llfan-ustc/huang2022/blob/main/catalog.csv}.

\begin{longrotatetable}
\begin{deluxetable*}{ccccCCCCCC}
\tablecaption{Properties of 2649 massive ETGs. (first 15 rows are shown)\label{tab:tidal}}
\tablewidth{\textwidth}
\decimalcolnumbers
\tabletypesize{\scriptsize}
\tablehead{SDSS\_NAME & R.A.(deg) & Dec.(deg) & redshift & \log M_*/M_{\odot} & r \rm (mag) & i \rm (mag) & r_{\rm tidal} \rm (mag) & i_{\rm tidal} \rm (mag) & \log f_{\rm tidal}}
\startdata
J000006.67+003016.7 & 0.02780 & 0.50465 & 0.109 & 11.03 \pm 0.13 & 16.957 \pm 0.006 & 16.547 \pm 0.005 & 22.40 \pm 0.10 & 22.06 \pm 0.09 & -2.19 \pm 0.21 \\
J000051.56+005323.8 & 0.21485 & 0.88998 & 0.104 & 11.16 \pm 0.12 & 16.410 \pm 0.005 & 16.032 \pm 0.004 & 22.02 \pm 0.08 & 21.65 \pm 0.08 & -2.25 \pm 0.21 \\
J000238.70+002309.8 & 0.66126 & 0.38611 & 0.088 & 11.31 \pm 0.15 & 15.952 \pm 0.004 & 15.488 \pm 0.003 & 20.06 \pm 0.03 & 20.37 \pm 0.04 & -1.78 \pm 0.20 \\
J000239.89+010040.7 & 0.66625 & 1.01129 & 0.087 & 11.01 \pm 0.14 & 16.421 \pm 0.005 & 16.021 \pm 0.004 & 24.51 \pm 0.25 & 99.00 \pm 99.00 & -3.54 \pm 0.22 \\
J000243.08+003419.2 & 0.67953 & 0.57202 & 0.078 & 11.08 \pm 0.12 & 16.182 \pm 0.004 & 15.750 \pm 0.004 & 22.05 \pm 0.08 & 21.65 \pm 0.07 & -2.36 \pm 0.21 \\
J000322.48+010328.3 & 0.84370 & 1.05789 & 0.079 & 11.68 \pm 0.10 & 14.842 \pm 0.002 & 14.399 \pm 0.002 & 19.54 \pm 0.02 & 19.80 \pm 0.03 & -2.00 \pm 0.21 \\
J000406.41+003303.5 & 1.02674 & 0.55101 & 0.086 & 11.33 \pm 0.12 & 15.904 \pm 0.004 & 15.424 \pm 0.003 & 21.95 \pm 0.08 & 22.51 \pm 0.12 & -2.58 \pm 0.22 \\
J000449.18+002800.9 & 1.20496 & 0.46698 & 0.095 & 11.20 \pm 0.09 & 16.443 \pm 0.005 & 15.925 \pm 0.005 & 22.39 \pm 0.11 & 22.36 \pm 0.29 & -2.47 \pm 0.22 \\
J000458.59+005743.4 & 1.24415 & 0.96210 & 0.095 & 11.48 \pm 0.11 & 15.737 \pm 0.004 & 15.306 \pm 0.003 & 19.88 \pm 0.03 & 19.45 \pm 0.02 & -1.66 \pm 0.19 \\
J000510.50+000540.4 & 1.29381 & 0.09458 & 0.138 & 11.18 \pm 0.14 & 17.013 \pm 0.006 & 16.534 \pm 0.005 & 23.04 \pm 0.13 & 22.88 \pm 0.14 & -2.47 \pm 0.22 \\
J000512.06-000902.6 & 1.30031 & -0.15073 & 0.139 & 11.44 \pm 0.10 & 16.483 \pm 0.005 & 16.073 \pm 0.004 & 20.94 \pm 0.05 & 20.49 \pm 0.04 & -1.78 \pm 0.20 \\
J000602.45-002214.2 & 1.51024 & -0.37061 & 0.114 & 11.01 \pm 0.14 & 17.087 \pm 0.007 & 16.657 \pm 0.005 & 24.38 \pm 0.24 & 23.51 \pm 0.18 & -2.82 \pm 0.22 \\
J000626.21-011110.8 & 1.60922 & -1.18634 & 0.096 & 11.30 \pm 0.13 & 16.552 \pm 0.005 & 16.037 \pm 0.004 & 22.87 \pm 0.14 & 22.28 \pm 0.12 & -2.51 \pm 0.22 \\
J000629.41-011213.8 & 1.62255 & -1.20385 & 0.098 & 11.08 \pm 0.11 & 16.901 \pm 0.006 & 16.399 \pm 0.005 & 22.50 \pm 0.11 & 21.59 \pm 0.07 & -2.15 \pm 0.21 \\
\enddata
\tablecomments{Columns: (1) the SDSS name. (2) right ascension from HSC-SSP PDR3. (3) declination from HSC-SSP PDR3. (4) redshift from SDSS DR16. (5) stellar mass from SDSS DR16 using the method of \citet{Chen2012}. (6)(7) magnitude of ETGs. (8)(9) magnitude of tidal features. (10) mean flux fraction of tidal features of $r$-band and $i$-band. Columns (6)--(10) are results given by this paper. Random noise and gain of the instrument are considered in the errors shown in columns (6)--(9). The error of $f_{\rm tidal}$ is calculated with the fitting function $\Delta f_{\rm tidal}=-1.40f_{\rm tidal}+0.22$ [dex], which is estimated from the scatter of $f_{\rm tidal}$ in $r$-band and $i$-band. Notice that the data is noisy for $f_{\rm tidal}<1\%$, especially for $f_{\rm tidal}\la 0.5\%$.}
\end{deluxetable*}
\end{longrotatetable}

\section{The Maximum A Posterior Estimation} \label{sec:mae}
In this section, we describe the method used to estimate the slope $\alpha$ of the exponential distribution of $f_{\rm tidal}~(f_{\rm tidal}\geq0.01)$ in Section \ref{subsubsec:fluxfrac}.

According to the Bayes theorem, if the given dataset $\vec{X}=(X_1, ..., X_n)$ is drawn from a distribution with an unknown parameter $\alpha$, the posterior probability of $\alpha$ satisfies: 
\begin{equation}
    {\rm P}(\alpha|\vec{X})\propto{\rm P}(\vec{X}|\alpha){\rm P}(\alpha)=\prod_{i=1}^{n}{\rm P}(X_i|\alpha){\rm P}(\alpha),\label{eq:b1}
\end{equation}
where ${\rm P}(\alpha)$ is the prior probability of $\alpha$. The equality holds because the data $X_i$'s are independent of each other. Maximizing the last quantity in Equation \ref{eq:b1} over a range of the parameter  $\alpha$ then gives an estimate of $\alpha$. In our work, we adapt a uniform prior distribution ${\rm P}(\alpha)$ that spans $50\leq\alpha\leq160$. Given the observed values of $f_{\rm tidal}$'s, the posterior probability of the slope $\alpha$ is proportional to:

\begin{equation}
    {\rm P}(\alpha|f_{\rm tidal})\propto\prod_{f_{\rm tidal,i}\geq0.01}\frac{1}{\alpha}\exp\left[-\alpha(f_{\rm tidal,i}-0.01)\right].
\end{equation}
The errors of $\alpha$ in Section \ref{sec:measure} are given by the 16th and the 84th percentiles of its posterior probability distribution.

%





\bibliography{Tidal_Feature_ETGs}{}
\bibliographystyle{aasjournal}



\end{CJK*}
\end{document}